\documentclass{IEEEtran}
\usepackage[T1]{fontenc}
\usepackage[latin9]{inputenc}
\usepackage{calc}
\usepackage{amsmath}
\usepackage{amsthm}
\usepackage{amssymb}
\usepackage{graphicx}
\usepackage[unicode=true,
 bookmarks=true,bookmarksnumbered=false,bookmarksopen=false,
 breaklinks=false,pdfborder={0 0 1},backref=false,colorlinks=false]
 {hyperref}
\hypersetup{pdftitle={Your Title},
 pdfauthor={Your Name},
 pdfpagelayout=OneColumn,pdfnewwindow=true,pdfstartview=XYZ,plainpages=false}

\makeatletter
\theoremstyle{plain}
\newtheorem{thm}{\protect\theoremname}
\theoremstyle{plain}
\newtheorem{prop}[thm]{\protect\propositionname}
\theoremstyle{remark}
\newtheorem{rem}[thm]{\protect\remarkname}
\theoremstyle{plain}
\newtheorem{lem}[thm]{\protect\lemmaname}
\theoremstyle{plain}
\newtheorem{cor}[thm]{\protect\corollaryname}

\usepackage{caption}
\usepackage{algorithm}
\usepackage{algpseudocode}
\usepackage{algorithmicx}
\usepackage{array}
\usepackage{boldline}
\usepackage{makecell}
\usepackage{booktabs}
\usepackage{multirow}
\usepackage{threeparttable}
\newcounter{mylabelcounter}
\newcommand{\labelText}[2]{%
#1\refstepcounter{mylabelcounter}%
\immediate\write\@auxout{%
  \string\newlabel{#2}{{1}{\thepage}{{\unexpanded{#1}}}{mylabelcounter.\number\value{mylabelcounter}}{}}%
}%
}

\providecommand{\lemmaname}{Lemma}
\providecommand{\propositionname}{Proposition}
\providecommand{\remarkname}{Remark}
\providecommand{\theoremname}{Theorem}

\@ifundefined{showcaptionsetup}{}{%
 \PassOptionsToPackage{caption=false}{subfig}}
\usepackage{subfig}
\makeatother

\providecommand{\corollaryname}{Corollary}
\providecommand{\lemmaname}{Lemma}
\providecommand{\propositionname}{Proposition}
\providecommand{\remarkname}{Remark}
\providecommand{\theoremname}{Theorem}

\begin{document}
\title{Discerning and Enhancing the Weighted Sum-Rate Maximization Algorithms
in Communications}
\author{Zepeng Zhang,~\IEEEmembership{Student Member,~IEEE}, Ziping Zhao,~\IEEEmembership{Member,~IEEE},
Kaiming Shen,~\IEEEmembership{Member,~IEEE},\\
 Daniel P. Palomar,~\IEEEmembership{Fellow,~IEEE}, and Wei Yu,~\IEEEmembership{Fellow,~IEEE}\thanks{This paper was presented in part at the 48th IEEE International Conference
on Acoustics, Speech, and Signal Processing (ICASSP), Rhodes Island,
Greece, June 4th--10th, 2023 \cite{zhang2023enhancing}.}}
\maketitle
\begin{abstract}
Weighted sum-rate (WSR) maximization plays a critical role in communication
system design. This paper examines three optimization methods for
WSR maximization, which ensure convergence to stationary points: two
block coordinate ascent (BCA) algorithms, namely, weighted sum-minimum
mean-square error (WMMSE) and WSR maximization via fractional programming
(WSR-FP), along with a minorization-maximization (MM) algorithm, WSR
maximization via MM (WSR-MM). Our contributions are threefold. Firstly,
we delineate the exact relationships among WMMSE, WSR-FP, and WSR-MM,
which, despite their extensive use in the literature, lack a comprehensive
comparative study. By probing the theoretical underpinnings linking
the BCA and MM algorithmic frameworks, we reveal the direct correlations
between the equivalent transformation techniques, essential to the
development of WMMSE and WSR-FP, and the surrogate functions pivotal
to WSR-MM. Secondly, we propose a novel algorithm, WSR-MM+, harnessing
the flexibility of selecting surrogate functions in MM framework.
By circumventing the repeated matrix inversions in the search for
optimal Lagrange multipliers in existing algorithms, WSR-MM+ significantly
reduces the computational load per iteration and accelerates convergence.
Thirdly, we reconceptualize WSR-MM+ within the BCA framework, introducing
a new equivalent transform, which gives rise to an enhanced version
of WSR-FP, named as WSR-FP+. We further demonstrate that WSR-MM+ can
be construed as the basic gradient projection method. This perspective
yields a deeper understanding into its computational intricacies.
Numerical simulations corroborate the connections between WMMSE, WSR-FP,
and WSR-MM and confirm the efficacy of the proposed WSR-MM+ and WSR-FP+
algorithms.
\end{abstract}

\begin{IEEEkeywords}
Rate maximization, weighted sum-minimum mean-square error (WMMSE),
fractional programming (FP), block coordinate ascent (BCA), minorization-maximization
(MM), equivalent transform, surrogate function. 
\end{IEEEkeywords}

\section{Introduction}

Weighted sum-rate (WSR) maximization plays a central role in a wide
variety of communication system design tasks \cite{weeraddana2012weighted},
a typical case of which is to find the optimal beamforming vectors
for the multi-antenna channels \cite{christensen2008weighted,shi2011iteratively,shen2018fractional1,zhang2021weighted}.
This problem is proved to be NP-hard \cite{luo2008dynamic,liu2010coordinated};
the state-of-the-art for WSR maximization is generally to attain a
stationary point solution via iterative algorithms, notably including
the weighted sum-minimum mean-square error (WMMSE) algorithms \cite{christensen2008weighted,shi2011iteratively},
the WSR maximization via fractional programming (WSR-FP) algorithms
\cite{shen2018fractional1,shen2019optimization}, and the WSR maximization
via minorization-maximization (WSR-MM) algorithms \cite{zhang2021weighted},
each offering a distinct strategy for WSR beamforming problems.

As three common non-convex optimization approaches, WMMSE, WSR-FP,
and WSR-MM are driven by completely different motivations. The appeal
of WMMSE originates from a foundational principle in signal processing:
maximizing the signal-to-interference-plus-noise ratio (SINR) equates
to minimizing the mean squared error (MSE) of the received signal.
The idea of casting the WSR maximization problem into the weighted
sum-MSE minimization problem to facilitate the problem solving was
first proposed in \cite{christensen2008weighted} for the multiple-input
single-out (MISO) channels and later extended by \cite{shi2011iteratively}
to the multiple-input multiple-output (MIMO) case. A recent progress
in WMMSE aims to reduce the computational complexity by using the
range space of channels \cite{zhao2022rethinking}. In contrast, fractional
programming approaches work towards a mathematical goal---it reformulates
a fractional optimization problem with one or more ratios so as to
decouple every ratio term. The classic methods, i.e., Dinkelbach's
algorithm and Charnes-Cooper algorithm \cite{stancu2012fractional},
have achieved this goal for a single ratio. However, the use of the
classic fractional programming techniques in communication system
design was typically restricted to the single-ratio problem scenario
such as the efficiency maximization \cite{cheung2013achieving,zappone2015energy},
while the WSR maximization, a logarithmic fractional programming problem,
does not fall in this category. A new multi-ratio ``equivalent transform''
technique, known as the quadratic transform, was developed in \cite{shen2018fractional1}
to coordinate multiple links (with multiple SINRs) in wireless networks.
The capability of dealing with multiple ratios enables the extensive
applications of the quadratic transform in communication system design,
e.g., \cite{khan2020optimizing,shen2020enhanced,park2021collaborative}.
In particular, aided by the quadratic transform and the Lagrangian
dual transform for dealing with logarithmic functions, WSR-FP was
introduced for WSR maximization \cite{shen2018fractional1}. As for
the minorization-maximization (MM) method \cite{hunter2004tutorial,sun2016majorization},
it involves two steps: In the minorization step, a ``surrogate function''
that lower bounds the objective function is constructed; subsequently,
in the maximization step, this surrogate function is maximized. The
inherent flexibility in formulating surrogate functions allows for
MM algorithms that are tailored to exploit the specific nuances of
given problems. The MM-based method, WSR-MM \cite{zhang2021weighted},
was developed specifically to tackle the WSR maximization problems.

The trio of methodologies---WMMSE, WSR-FP, and WSR-MM---diverges
in foundation and approach. Both WSR-FP \cite{shen2018fractional1,shen2018fractional2}
and WSR-MM \cite{zhang2021weighted} originate from a mathematical
algorithmic perspective. In contrast, WMMSE \cite{christensen2008weighted,shi2011iteratively},
while later framed through a mathematical lens \cite{shi2015secure},
is anchored in the physical background of communications, relying
on transforming the WSR problem into a weighted sum-MSE minimization
problem. The WSR-MM method addresses the optimization problem by iteratively
constructing a surrogate function of the original objective, which
is then optimized in each iteration. In contrast, the WMMSE and WSR-FP
methods deploy equivalent transformations to the initial optimization
problem. They inadvertently increase the dimensionality of the optimization
variables by introducing auxiliary variables, thus leading to multi-block
optimization problems. In these scenarios, both the beamforming variables
and the auxiliary variables are optimized using block coordinate ascent
(BCA) techniques \cite{bertsekas1999nonlinear}. Noteworthy is the
non-unique decouple schemes when applying equivalent transforms in
WSR-FP and the non-unique variable block selection rules when applying
BCA, which makes them not two individual algorithms but two algorithm
categories. In this paper, we also provide novel insights and perspectives
on the WSR-FP algorithms, presenting findings previously unexplored
in the literature. 

Although birthed from distinct technical lineages, there exists a
strong connection among the WMMSE, WSR-FP, and WSR-MM algorithms.
While strides have been made to elucidate the interrelationships among
them \cite{BSUM,shen2018fractional2,shen2019optimization}, certain
elements remain sufficiently opaque. This paper endeavors to clarify
these connections through an exhaustive examination through two representative
communication systems: a MISO system with broadcast channels and a
MIMO system with interference channels. We demonstrate that WSR-FP
\cite{shen2018fractional1,shen2018fractional2} subsumes WMMSE \cite{christensen2008weighted,shi2011iteratively}
which, in turn, subsumes WSR-MM \cite{zhang2021weighted}. Additionally,
we establish that not all instances of WSR-FP align with the MM methodology;
rather, this interpretation is contingent upon the fulfillment of
specific criteria. We also elucidate the parallelism between the equivalent
transformations employed in WMMSE and WSR-FP and the intricate surrogate
function construction methods pivotal in the WSR-MM strategy. By taking
the minorization step in the WSR-MM method as a bifurcated surrogate
function development procedure \cite{zhang2021weighted}, we observe
that these two steps have one-to-one correspondences to the two equivalent
transforms adopted in WMMSE and WSR-FP. By employing specific variable
update orders in WMMSE and particular decoupling schemes alongside
variable update orders in WSR-FP, the updates of auxiliary variables
in both algorithms can be viewed as the construction of surrogate
functions within WSR-MM. It is important to note that, unlike the
methodologies in \cite{shen2019optimization}, which retrospectively
associate certain WSR-FP algorithms with the MM algorithmic framework,
our analysis employs a forward-looking, constructive approach. This
proactive method enhances our understanding of existing algorithms
and inspires the development of novel ones for WSR maximization, fully
leveraging the capabilities of the MM framework.

In the problem of WSR maximization subject to the total power budget
constraint, a particularly nuanced aspect of the prevailing WMMSE,
WSR-FP, and WSR-MM approaches is the calibration process for determining
the optimal Lagrange multiplier that satisfies the power constraint.
This process is computationally intensive due to the recurrent requirement
for calculating matrix (pseudo-)inversions. To forge a more efficient
algorithm with lower per-iteration computational complexity, we incorporate
an extra minorization phase in the application of the MM technique
to the WSR maximization problem. This leads to the development of
a novel single-loop algorithm, termed WSR-MM+, which offers analytical
solutions and eliminates the need for adjusting the Lagrange multiplier.
Elaborating on this, we also interpret WSR-MM+ through the lens of
BCA, which paves the way for an inventive equivalent transform methodology.
Capitalizing on this technique, we introduce a novel class of single-loop
algorithms, designated as WSR-FP+. In addition, we establish that
WSR-MM+ can be construed as a projected gradient ascent method, which
inherently incorporates an adaptive step size. This novel conceptualization
not only streamlines the computational mechanisms but also sets the
stage for further algorithmic developments.

To make it clear, main contributions of this paper are summarized
as follows:
\begin{itemize}
\item A thorough and systematic examination of the interconnections between
WMMSE, WSR-FP, and WSR-MM algorithms is presented. It is also established
that the equivalent transformations deployed in formulating WMMSE
and WSR-FP are intrinsically related to the surrogate function construction
techniques utilized in WSR-MM. 
\item To spare the Lagrange multiplier tuning in existing iterative WSR
maximization algorithms, we introduce two novel single-loop algorithms:
the MM-inspired WSR-MM+ and its BCA counterpart WSR-FP+.
\item The proposed WSR-MM+ is characterized as a projected gradient ascent
method with an implicit step size.
\item Complexity analyses of existing and proposed algorithms are conducted.
The relationships between WMMSE, WSR-FP, and WSR-MM, as well as the
superior performances of WSR-MM+ and WSR-FP+, are substantiated through
experiments conducted in both MISO and MIMO system contexts.
\end{itemize}
The rest of this paper is organized as follows. Section \ref{sec:BCA-MM}
starts with a brief introduction to the BCA and MM frameworks. To
more easily grasp the concepts underlying the WMMSE, WSR-FP, and WSR-MM
methods, we will first review their applications in the MISO case
in Section \ref{Section: MISO} before progressing to the MIMO scenario
in Section \ref{Section: MIMO}. The intricate relationships among
the WMMSE, WSR-FP, and WSR-MM methods are dissected in Section \ref{Section: connection}.
The proposed WSR-MM+ algorithm, along with its interpretations through
BCA and projected gradient ascent, is comprehensively introduced in
Section \ref{Section: analytical solution}. Algorithm complexity
comparisons are given in Section \ref{Section: Complexity}. Finally,
Section \ref{Section: Experiments} provides simulation results, followed
by conclusions and discussions in Section \ref{Section: Conclusion}.

\textit{Notation:} Italic letters, boldface lower-case letters, and
boldface upper-case letters denote scalars, column vectors, and matrices,
respectively. We denote by $\mathbf{I}$ the identity matrices. The
real and complex numbers are denoted by $\mathbb{R}$ and $\mathbb{C}$,
respectively. $N$-dimensional complex vectors are denoted by $\mathbb{C}^{N}$.
$N\times M$-dimensional complex matrices and Hermitian matrices are
denoted by $\mathbb{C}^{N\times M}$ and $\mathbb{H}^{N}$, respectively.
Given $\mathbf{A}$, $\mathbf{B}\in\mathbb{H}^{N}$, $\mathbf{A}\succ\mathbf{B}$
and $\mathbf{A}\succeq\mathbf{B}$ stand for $\mathbf{A}-\mathbf{B}$
is positive semidefinite and positive definite respectively. $(\cdot)^{*}$,
$(\cdot)^{\mathsf{H}}$, $(\cdot)^{-1}$, and $(\cdot)^{\dagger}$
are used to denote the matrix conjugate, Hermitian, inverse, and pseudo-inverse,
respectively. $\mathrm{Re}(\cdot)$ denotes the real part of a scalar.
$\left\Vert \cdot\right\Vert $ denotes the vector $2$-norm or the
matrix Frobenius norm. $\mathrm{det}(\cdot)$, $\mathrm{tr}(\cdot)$,
and $\lambda_{\max}(\cdot)$ denote the determinant, trace, and largest
eigenvalue of a matrix, respectively. $\frac{\partial f\left(x\right)}{\partial x}$
represents the partial derivative of function $f$ at $x$. $f(n)={\cal O}(g(n))$
means there are positive constants $c$ and $k$, such that $0\leq f(n)\leq cg(n)$
for all $n\geq k$.

\section{Block Coordinate Ascent and Minorization-Maximization \label{sec:BCA-MM}}

We provide a brief overview of two classical iterative optimization
methods---block coordinate ascent (BCA) \cite{bertsekas1999nonlinear}
and minorization-maximization (MM) \cite{hunter2004tutorial,sun2016majorization}---for
solving a maximization problem. To illustrate the fundamental concept
of BCA, we examine a problem with three variables:
\begin{equation}
\begin{aligned} & \underset{u,v,x}{\text{maximize}} &  & g(u,v,x)\\
 & \text{subject\ to} &  & u\in\mathcal{U},v\in\mathcal{V},x\in\mathcal{X},
\end{aligned}
\label{eq:BCA}
\end{equation}
where $\mathcal{U}$, $\mathcal{V}$, and $\mathcal{X}$ denote the
feasible domains for $u$, $v$, and $x$, respectively. Given initial
feasible iterates, BCA iteratively refines one block of variables
at a time, holding the others constant. This process is described
by the following update rules for $t=0,1,2,\ldots$:
\begin{equation}
\begin{cases}
u^{(t+1)}\in\arg\max_{u\in\mathcal{U}}g(u,v^{(t)},x^{(t)})\\
v^{(t+1)}\in\arg\max_{v\in\mathcal{V}}g(u^{(t+1)},v,x^{(t)})\\
x^{(t+1)}\in\arg\max_{x\in\mathcal{X}}g(u^{(t+1)},v^{(t+1)},x).
\end{cases}\label{eq:BCA update}
\end{equation}
The three variable blocks can be updated following a cyclic pattern
or through other block selection rules \cite{BSUM,phan2023inertial}.
Turning to MM, we consider the following problem: 
\begin{equation}
\begin{aligned} & \underset{x}{\text{maximize}} &  & f(x)\\
 & \text{subject\ to} &  & x\in\mathcal{X},
\end{aligned}
\label{eq:MM}
\end{equation}
where ${\cal X}$ denotes the feasible set for $x$. MM, starting
from a feasible initial point, tackles this problem by solving a sequence
of surrogate optimization problems. Specifically, at $x^{(t)}$, the
MM update rule is:
\begin{equation}
\mathbf{x}^{(t+1)}\in\arg\max_{x\in\mathcal{X}}\ell(x,x^{(t)}),\label{eq:MM update}
\end{equation}
where $\ell(x,x^{(t)})$ represents the surrogate function of $f(x)$
at $x^{(t)}$ and must satisfy the conditions $\ell(x,x^{(t)})\leq f(x)$
and $\ell(x^{(t)},x^{(t)})=f(x^{(t)})$ for all $x$ and $x^{(t)}$
within the set $\mathcal{X}$. For discussions regarding the convergence
properties of BCA and MM, interested readers are directed to \cite{grippo2000convergence,tseng2001convergence,BSUM,phan2023inertial}.

\section{WSR Maximization Over MISO Broadcast Channel \label{Section: MISO}}

\subsection{System Model and Problem Formulation}

Consider a communication system of downlink transmission, where $K$
single-antenna users is served by a base station with $M$ antennas.
For each user $k$ $(k=1,\ldots,K)$, we define $\mathbf{w}_{k}\in\mathbb{C}^{M}$
as the transmit beamforming vector and $\mathbf{h}_{k}\in\mathbb{C}^{M}$
as the channel from the base station to the user. The signal received
at the $k$-th user, $y_{k}$, can be described as follows:
\begin{equation}
y_{k}=\mathbf{h}_{k}^{\mathsf{H}}\mathbf{w}_{k}s_{k}+\sum_{j=1,j\neq k}^{K}\mathbf{h}_{k}^{\mathsf{H}}\mathbf{w}_{j}s_{j}+n_{k},
\end{equation}
where $s_{k}$ denotes the symbol intended for user $k$ with zero
mean and unit variance and $n_{k}\sim\mathcal{CN}(0,\sigma_{k}^{2})$
is the additive white Gaussian noise. The SINR for the $k$-th user
is given by\footnote{Throughout this paper, the set $\{\mathbf{w}_{i}\}$ is used to collectively
represent the beamforming vectors $\mathbf{w}_{1},\ldots,\mathbf{w}_{K}$,
and similar notations are adopted for other parameter sets.} 
\begin{equation}
\mathsf{SINR}_{k}(\{\mathbf{w}_{i}\})=\frac{\bigl|\mathbf{h}_{k}^{\mathsf{H}}\mathbf{w}_{k}\bigr|^{2}}{\sum_{j=1,j\neq k}^{K}\bigl|\mathbf{h}_{k}^{\mathsf{H}}\mathbf{w}_{j}\bigr|^{2}+\sigma_{k}^{2}},
\end{equation}
and the corresponding rate can be calculated as
\begin{equation}
\mathsf{R}_{k}(\{\mathbf{w}_{i}\})=\log\left(1+\mathsf{SINR}_{k}(\{\mathbf{w}_{i}\})\right).
\end{equation}
Our objective is to maximize the overall WSR of the system by optimizing
the beamformer vectors, i.e., 
\begin{equation}
\begin{aligned} & \underset{\{\mathbf{w}_{i}\}\in\mathcal{W}}{\text{maximize}} &  & f\left(\{\mathbf{w}_{i}\}\right)=\sum_{k=1}^{K}\omega_{k}\mathsf{R}_{k}(\{\mathbf{w}_{i}\}),\end{aligned}
\label{0 WSR Maximization}
\end{equation}
where $\omega_{k}$ is a weighting coefficient reflecting the relative
priority of user $k$ in the system and the beamformer is subject
to the following transmit power limit constraint:
\begin{equation}
\mathcal{W}=\Bigl\{\{\mathbf{w}_{i}\}\mid\sum_{k=1}^{K}\left\Vert \mathbf{w}_{k}\right\Vert ^{2}\leq P\Bigr\},
\end{equation}
with $P$ the total available transmit power at the base station.

The WSR maximization problem \eqref{0 WSR Maximization} is non-convex.
In the following, we will first briefly review three prevalent methods
for problem solving in the literature, namely, WMMSE \cite{shi2011iteratively},
WSR-FP \cite{shen2018fractional1}, and WSR-MM \cite{zhang2021weighted}.

\subsection{WMMSE Algorithms \label{Section: WMMSE for WSRMax}}

Following the WMMSE method proposed in \cite{christensen2008weighted,shi2011iteratively},
the WSR maximization problem \eqref{0 WSR Maximization} is recast
into an equivalent weighted sum-MSE minimization problem. Under the
independence assumption of $s_{k}$'s and $n_{k}$'s, the MSE between
the estimated signal and the original signal for user $k$ is given
by
\begin{equation}
\begin{aligned}e_{k}= & \mathbb{E}\bigl[\bigl|l_{k}^{*}(\mathbf{h}_{k}^{\mathsf{H}}\mathbf{w}_{k}s_{k}+\sum_{j=1,j\neq k}^{K}\mathbf{h}_{k}^{\mathsf{H}}\mathbf{w}_{j}s_{j}+n_{k})-s_{k}\bigr|^{2}\bigr]\\
= & \left|l_{k}^{*}\mathbf{h}_{k}^{\mathsf{H}}\mathbf{w}_{k}-1\right|^{2}+\sum_{j=1,j\neq k}^{K}\left|l_{k}^{*}\mathbf{h}_{k}^{\mathsf{H}}\mathbf{w}_{j}\right|^{2}+\sigma_{k}^{2}\left|l_{k}^{*}\right|^{2},
\end{aligned}
\end{equation}
where $l_{k}$ denotes the receive beamformer for user $k$.
\begin{prop}[\cite{shi2011iteratively}]
\label{Thm: WMMSE} Let $m_{1},\ldots,m_{K}\geq0$ be the weights.
The WSR maximization problem \eqref{0 WSR Maximization} is equivalent
to the following weighted sum-MSE minimization problem 
\begin{equation}
\begin{aligned} & \underset{\{l_{i}\},\{m_{i}\},\{\mathbf{w}_{i}\}\in\mathcal{W}}{\text{maximize}} &  & \sum_{k=1}^{K}\omega_{k}\left(\log m_{k}-m_{k}e_{k}\right),\end{aligned}
\label{WMMSE: Weight Sum-MSE}
\end{equation}
in the sense that they attain the identical optimal solution. 
\end{prop}
Based on Proposition \ref{Thm: WMMSE}, Problem \eqref{0 WSR Maximization}
can be addressed via solving Problem \eqref{WMMSE: Weight Sum-MSE},
a three-block optimization problem with beamforming variables $\{\mathbf{w}_{i}\}$
and auxiliary variables $\{l_{i}\}$ and $\{m_{i}\}$. Solving Problem
\eqref{WMMSE: Weight Sum-MSE} based on BCA leads to the WMMSE algorithms
\cite{christensen2008weighted,shi2011iteratively}. Given initial
feasible iterates, \nameref{label:MISO-WMMSE} summarizes the variable
update steps in each round.\footnote{In this paper, instead of using the iteration index, we use $\underline{\mathbf{x}}$
to denote a variable $\mathbf{x}$ with most recently updated value.}

\noindent %
\noindent\fbox{\begin{minipage}[t]{1\columnwidth - 2\fboxsep - 2\fboxrule}%
\begin{center}
\labelText{\textbf{A1}}{label:MISO-WMMSE}: A WMMSE algorithm for
MISO beamforming \cite{shi2011iteratively}
\begin{align*}
\text{S1: } & l_{k}=\frac{\mathbf{h}_{k}^{\mathsf{H}}\underline{\mathbf{w}}_{k}}{\sum_{j=1}^{K}\left|\mathbf{h}_{k}^{\mathsf{H}}\underline{\mathbf{w}}_{j}\right|^{2}+\sigma_{k}^{2}},\\
\text{S2: } & m_{k}=1+\mathsf{SINR}_{k}(\{\underline{\mathbf{w}}_{i}\}),\\
\text{S3: } & \mathbf{w}_{k}=\bigl(\sum_{j=1}^{K}\omega_{j}\underline{m}_{j}\left|\underline{l}_{j}\right|^{2}\mathbf{h}_{j}\mathbf{h}_{j}^{\mathsf{H}}+\mu\mathbf{I}\bigr)^{\dagger}\omega_{k}\underline{m}_{k}\underline{l}_{k}\mathbf{h}_{k}.
\end{align*}
\par\end{center}%
\end{minipage}}

Due to the block-wise convex nature of Problem \eqref{WMMSE: Weight Sum-MSE},
the subproblems pertaining to $\{l_{i}\}$ and $\{m_{i}\}$ are efficiently
resolved using their respective first-order optimality conditions,
whereas the subproblem related to $\{\mathbf{w}_{i}\}$, which forms
a quadratic constrained quadratic program, is solved via Lagrangian
multiplier method. In the update of $\mathbf{w}_{k}$, we have employed
the pseudo-inverse to generally address the scenario where the matrix
$\sum_{j=1}^{K}\omega_{j}\underline{m}_{j}\left|\underline{l}_{j}\right|^{2}\mathbf{h}_{j}\mathbf{h}_{j}^{\mathsf{H}}$
is of low rank, a particular case which has not been thoroughly considered
in the WSR maximization literature; $\mu$ is the Lagrangian multiplier
for the power constraint, ascertained by
\begin{equation}
\mu=\min\Bigl\{\mu\geq0:\sum_{k=1}^{K}\left\Vert \mathbf{w}_{k}\left(\mu\right)\right\Vert ^{2}\leq P\Bigr\},\label{eq:bisection mu MISO}
\end{equation}
which can be determined via one dimensional search \cite{bertsekas1999nonlinear}. 
\begin{rem}
The variable update order in WMMSE is not unique. An alternative algorithm
to \nameref{label:MISO-WMMSE} is given as follows:
\begin{equation}
\dashrightarrow\underbrace{\text{\nameref{label:MISO-WMMSE}-S2}\rightarrow\text{\nameref{label:MISO-WMMSE}-S1}\rightarrow\text{\nameref{label:MISO-WMMSE}-S3}}_{\text{one round of variable update}}\rightarrow\text{\nameref{label:MISO-WMMSE}-S2}\dashrightarrow.\label{eq:WMMSE another update order}
\end{equation}
\end{rem}

\subsection{WSR-FP Algorithms \label{Section: FP for WSRMax}}

The WSR-FP algorithm, as described in \cite{shen2018fractional1},
employs a similar strategy to WMMSE by recasting \eqref{0 WSR Maximization}
into a multi-block optimization problem. This is achieved through
two transformative steps: the Lagrangian dual transform and the quadratic
transform. Following these transformations, the BCA method is applied
to iteratively solve the resulting problem.
\begin{prop}[Lagrangian dual transform \cite{shen2018fractional2}]
\label{1 Transform: FP Lagrangian dual} Given ratios $\frac{c_{k}(\mathbf{x})}{d_{k}(\mathbf{x})}$
with $c_{k}(\mathbf{x})\geq0$ and $d_{k}(\mathbf{x})>0$ for $k=1,\ldots,K$,
the weighted sum-of-logarithmic-ratios maximization problem: 
\begin{equation}
\begin{aligned} & \underset{\mathbf{x}\in\mathcal{X}}{\text{maximize}} &  & \sum_{k=1}^{K}\omega_{k}\log\left(1+\frac{c_{k}(\mathbf{x})}{d_{k}(\mathbf{x})}\right)\end{aligned}
\label{FP: weighted sum-of-logrithms maximization}
\end{equation}
is equivalent to 
\begin{equation}
\begin{aligned} & \underset{\{\gamma_{i}\},\mathbf{x}\in\mathcal{X}}{\text{maximize}} &  & \sum_{k=1}^{K}\omega_{k}\left(\log\left(1+\gamma_{k}\right)-\gamma_{k}+\frac{\left(1+\gamma_{k}\right)c_{k}(\mathbf{x})}{n_{k}(\mathbf{x})+d_{k}(\mathbf{x})}\right),\end{aligned}
\label{FP: Lagrangian dual transformed problem}
\end{equation}
in the sense that they attain the identical optimal solution with
the identical optimal objective value. 
\end{prop}
\begin{prop}[Quadratic transform \cite{shen2018fractional1}]
\label{1 Transform: FP Quadratic Scalar} Given nondecreasing functions
$q_{k}(\cdot)$ and ratios $\frac{\left|c_{k}(\mathbf{x})\right|^{2}}{d_{k}(\mathbf{x})}$
with $c_{k}(\mathbf{x})\in\mathbb{C}$ and $d_{k}(\mathbf{x})>0$
for $k=1,\ldots,K$, the following problem: 
\begin{equation}
\begin{aligned} & \underset{\mathbf{x}\in\mathcal{X}}{\text{maximize}} &  & \sum_{k=1}^{K}q_{k}\left(\frac{\left|c_{k}(\mathbf{x})\right|^{2}}{d_{k}(\mathbf{x})}\right)\end{aligned}
\label{FP: sum-of-functions-of-ratio}
\end{equation}
is equivalent to 
\begin{equation}
\begin{aligned} & \underset{\{\phi_{i}\},\mathbf{x}\in\mathcal{X}}{\text{maximize}} &  & \sum_{k=1}^{K}q_{k}\left(2\mathrm{Re}\left(\phi_{k}^{*}c_{k}(\mathbf{x})\right)-\left|\phi_{k}\right|^{2}d_{k}(\mathbf{x})\right),\end{aligned}
\label{FP: Quadratic transformed problem}
\end{equation}
in the sense that they attain the identical optimal solution with
the identical optimal objective value. 
\end{prop}
Applying Proposition \ref{1 Transform: FP Lagrangian dual} to Problem
\eqref{0 WSR Maximization} and setting $c_{k}(\mathbf{x})=\bigl|\mathbf{h}_{k}^{\mathsf{H}}\mathbf{w}_{k}\bigr|^{2}$
and $d_{k}(\mathbf{x})=\sum_{j=1,j\neq k}^{K}\bigl|\mathbf{h}_{k}^{\mathsf{H}}\mathbf{w}_{j}\bigr|^{2}+\sigma_{k}^{2}$,
we obtain
\begin{equation}
\begin{aligned} & \underset{\{\gamma_{i}\},\{\mathbf{w}_{i}\}\in\mathcal{W}}{\text{maximize}} &  & \sum_{k=1}^{K}\omega_{k}\biggl(\log(1+\gamma_{k})-\gamma_{k}\\
 &  &  & \hspace{1.5cm}+\frac{(1+\gamma_{k})|\mathbf{h}_{k}^{\mathsf{H}}\mathbf{w}_{k}|^{2}}{\sum_{j=1}^{K}|\mathbf{h}_{k}^{\mathsf{H}}\mathbf{w}_{j}|^{2}+\sigma_{k}^{2}}\biggr).
\end{aligned}
\label{FP: WSR Lagrangian transform}
\end{equation}
The ratios in the objective of Problem \eqref{FP: WSR Lagrangian transform}
can be further decoupled using Proposition \ref{1 Transform: FP Quadratic Scalar}
with $q_{k}(x)=x$, $c_{k}(\mathbf{x})=\sqrt{\omega_{k}\left(1+\gamma_{k}\right)}\mathbf{h}_{k}^{\mathsf{H}}\mathbf{w}_{k}$,
and $d_{k}(\mathbf{x})=\sum_{j=1}^{K}\bigl|\mathbf{h}_{k}^{\mathsf{H}}\mathbf{w}_{j}\bigr|^{2}+\sigma_{k}^{2}$,
leading to the following problem: 
\begin{equation}
\begin{aligned} & \hspace{-0.2cm}\underset{\{\gamma_{i}\},\{\phi_{i}\},\{\mathbf{w}_{i}\}\in\mathcal{W}}{\text{maximize}} &  & \hspace{-0.2cm}\sum_{k=1}^{K}\Bigl(2\mathrm{Re}\bigl(\phi_{k}^{*}\sqrt{\omega_{k}\left(1+\gamma_{k}\right)}\mathbf{h}_{k}^{\mathsf{H}}\mathbf{w}_{k}\bigr)\\
 &  &  & \hspace{-2cm}-\left|\phi_{k}\right|^{2}\bigl(\sigma_{k}^{2}+\sum_{j=1}^{K}\bigl|\mathbf{h}_{k}^{\mathsf{H}}\mathbf{w}_{j}\bigr|^{2}\bigr)+\omega_{k}\bigl(\log\left(1+\gamma_{k}\right)-\gamma_{k}\bigr)\Bigr).
\end{aligned}
\label{FP: Final Problem}
\end{equation}
Problem \eqref{FP: Final Problem} is convex in each variable block---$\{\gamma_{i}\}$,
$\{\phi_{i}\}$, and $\{\mathbf{w}_{i}\}$, and can be solved by resorting
to the BCA method.

In \cite{shen2018fractional1}, an ``unconventional'' type of BCA
method was proposed, which is to update $\{\gamma_{i}\}$ by solving
Problem \eqref{FP: WSR Lagrangian transform} and to update $\{\phi_{i}\}$
and $\{\mathbf{w}_{i}\}$ by solving Problem \eqref{FP: Final Problem}.
The variable update steps are summarized in \nameref{label:MISO-FP}
with $\mu$ determined by \eqref{eq:bisection mu MISO}.

\noindent %
\noindent\fbox{\begin{minipage}[t]{1\columnwidth - 2\fboxsep - 2\fboxrule}%
\begin{center}
\labelText{\textbf{A2}}{label:MISO-FP}: A WSR-FP algorithm for
MISO beamforming \cite{shen2018fractional1}
\begin{align*}
\text{S1: } & \gamma_{k}=\mathsf{SINR}_{k}(\{\underline{\mathbf{w}}_{i}\}),\\
\text{S2: } & \phi_{k}=\frac{\sqrt{\omega_{k}\left(1+\underline{\gamma}_{k}\right)}\mathbf{h}_{k}^{\mathsf{H}}\underline{\mathbf{w}}_{k}}{\sum_{j=1}^{K}\bigl|\mathbf{h}_{k}^{\mathsf{H}}\underline{\mathbf{w}}_{j}\bigr|^{2}+\sigma_{k}^{2}},\\
\text{S3: } & \mathbf{w}_{k}=\Bigl(\sum_{j=1}^{K}|\underline{\phi}_{j}|^{2}\mathbf{h}_{j}\mathbf{h}_{j}^{\mathsf{H}}+\mu\mathbf{I}\Bigr)^{\dagger}\sqrt{\omega_{k}(1+\underline{\gamma}_{k})}\underline{\phi}_{k}\mathbf{h}_{k}.
\end{align*}
\par\end{center}%
\end{minipage}}

Despite deviating from the traditional BCA, i.e., solving all variable
blocks based on the single problem \eqref{FP: Final Problem}, convergence
of \nameref{label:MISO-FP} to stationary points is assured as demonstrated
in \cite{shen2018fractional2}. 
\begin{rem}
Employing the conventional BCA approach to addressing Problem \eqref{FP: Final Problem}
would result in identical update procedures for the $\{\phi_{i}\}$
and $\{\mathbf{w}_{i}\}$ variable blocks as delineated in \nameref{label:MISO-FP}.
However, the update formula for $\gamma_{k}$ is modified to be
\begin{equation}
\gamma_{k}=\frac{1}{2}\Biggl(\frac{\underline{\phi}_{k}^{2}|\mathbf{h}_{k}^{\mathsf{H}}\underline{\mathbf{w}}_{k}|^{2}}{\omega_{k}}+\sqrt{\Bigl(\frac{\underline{\phi}_{k}^{2}|\mathbf{h}_{k}^{\mathsf{H}}\underline{\mathbf{w}}_{k}|^{2}}{\omega_{k}}\Bigr)^{2}-4\frac{\underline{\phi}_{k}^{2}|\mathbf{h}_{k}^{\mathsf{H}}\underline{\mathbf{w}}_{k}|^{2}}{\omega_{k}}}\Biggr).\label{eq:conventional gamma}
\end{equation}
Then, we can derive two variant WSR-FP algorithms characterized by
the update sequences: 
\begin{equation}
\dashrightarrow\underbrace{\text{Eq. \eqref{eq:conventional gamma}}\rightarrow\text{\nameref{label:MISO-FP}-S2}\rightarrow\text{\nameref{label:MISO-FP}-S3}}_{\text{one round of variable update}}\rightarrow\text{Eq. \eqref{eq:conventional gamma}}\dashrightarrow\label{eq:conventional BCD-1}
\end{equation}
and 
\begin{equation}
\dashrightarrow\underbrace{\text{\nameref{label:MISO-FP}-S2}\rightarrow\text{Eq. \eqref{eq:conventional gamma}}\rightarrow\text{\nameref{label:MISO-FP}-S3}}_{\text{one round of variable update}}\rightarrow\text{\nameref{label:MISO-FP}-S2}\dashrightarrow.\label{eq:conventional BCD-2}
\end{equation}
For the aforementioned algorithms, it can be demonstrated that every
limit point of the sequence they generate is indeed a stationary point
of Problem \eqref{0 WSR Maximization}, a result validated by the
general BCA convergence theorem \cite[Proposition 5]{grippo2000convergence}. 
\end{rem}
\begin{rem}
Observe that if we substitute \nameref{label:MISO-FP}-S2 into \eqref{eq:conventional gamma},
we will obtain \nameref{label:MISO-FP}-S1. This indicates that the
unconventional BCA update strategy outlined in \nameref{label:MISO-FP}
can be construed as executing a conventional BCA, that is, 
\begin{equation}
\dashrightarrow\underbrace{\text{\nameref{label:MISO-FP}-S2}\rightarrow\text{Eq. \eqref{eq:conventional gamma}}\rightarrow\text{\nameref{label:MISO-FP}-S2}\rightarrow\text{\nameref{label:MISO-FP}-S3}}_{\text{one round of variable update}}\rightarrow\text{\nameref{label:MISO-FP}-S2}\dashrightarrow.\label{eq:unconventional as conventional}
\end{equation}
The algorithm in \eqref{eq:unconventional as conventional} can be
viewed as an augmentation of the ones in \eqref{eq:conventional BCD-1}
or \eqref{eq:conventional BCD-2}, enriched by an additional updating
step \nameref{label:MISO-FP}-S2. This modification constitutes a
BCA algorithm that updates variables following an essentially cyclic
rule.
\end{rem}
To derive Problem \eqref{FP: Final Problem}, we have utilized a particular
decoupling approach to dealing with the ratio terms in the objective
of Problem \eqref{FP: WSR Lagrangian transform}. However, alternative
decoupling techniques are also available as suggested in \cite{shen2019optimization}.
For instance, Proposition \ref{1 Transform: FP Quadratic Scalar}
can be applied to Problem \eqref{FP: WSR Lagrangian transform} by
setting $q_{k}(x)=\left(1+\gamma_{k}\right)x$, $c_{k}(\mathbf{x})=\sqrt{\omega_{k}}\mathbf{h}_{k}^{\mathsf{H}}\mathbf{w}_{k}$,
and $d_{k}(\mathbf{x})=\sum_{j=1}^{K}\bigl|\mathbf{h}_{k}^{\mathsf{H}}\mathbf{w}_{j}\bigr|^{2}+\sigma_{k}^{2}$,
which is the particular formulation to be employed in Section \ref{Section: FP for WSR Maximization MIMO}
for the design of beamformers in MIMO systems. Additionally, by applying
Proposition \ref{1 Transform: FP Quadratic Scalar} to Problem \eqref{FP: WSR Lagrangian transform}
with $q_{k}(x)=\omega_{k}\left(1+\gamma_{k}\right)x$, $c_{k}(\mathbf{x})=\mathbf{h}_{k}^{\mathsf{H}}\mathbf{w}_{k}$,
and $d_{k}(\mathbf{x})=\sum_{j=1}^{K}\bigl|\mathbf{h}_{k}^{\mathsf{H}}\mathbf{w}_{j}\bigr|^{2}+\sigma_{k}^{2}$,
we arrive at the well-known weighted sum-MSE minimization problem
\eqref{WMMSE: Weight Sum-MSE}. Overall, the WSR-FP framework exhibits
versatility; by selecting from various decoupling strategies and sequences
for updating variables, we can develop diverse variations of WSR-FP
algorithms.

\subsection{WSR-MM Algorithms \label{Section: MM for WSRMax}}

In this section, we introduce the WSR-MM approach proposed in \cite{zhang2021weighted}.
We first introduce a pertinent result pivotal for constructing MM
surrogate functions for Problem \eqref{0 WSR Maximization}. 
\begin{prop}[\cite{zhang2021weighted}]
\label{1 Transform: MM log+quadratic} Given $\log(1+\frac{\left|x\right|^{2}}{z})$
with $x\in\mathbb{C}$ and $z>0$, at $(\underline{x},\underline{z})$
we have
\begin{equation}
\begin{aligned}\log\bigl(1+\frac{|x|^{2}}{z}\bigr)\geq & \log\bigl(1+\frac{|\underline{x}|^{2}}{\underline{z}}\bigr)-\frac{|\underline{x}|^{2}}{\underline{z}}+2\mathrm{Re}\bigl(\frac{\underline{x}^{*}}{\underline{z}}x\bigr)\\
 & -\frac{|\underline{x}|^{2}}{\underline{z}(\underline{z}+|\underline{x}|^{2})}(z+|x|^{2}),
\end{aligned}
\end{equation}
where the equality is attained when $\left(x,z\right)=\left(\underline{x},\underline{z}\right)$. 
\end{prop}
By invoking Proposition \ref{1 Transform: MM log+quadratic} and applying
it to $f\left(\{\mathbf{w}_{i}\}\right)$, where $x=\mathbf{h}_{k}^{\mathsf{H}}\mathbf{w}_{k}$
and $z=\sum_{j=1,j\neq k}^{K}\bigl|\mathbf{h}_{k}^{\mathsf{H}}\mathbf{w}_{j}\bigr|^{2}+\sigma_{k}^{2}$,
we can construct a surrogate function at $\{\underline{\mathbf{w}}_{i}\}$
as follows:
\begin{equation}
\begin{aligned}\ell\left(\{\mathbf{w}_{i}\},\{\underline{\mathbf{w}}_{i}\}\right)= & \sum_{k=1}^{K}\omega_{k}\Bigl(-a_{k}\sum_{j=1}^{K}|\mathbf{h}_{k}^{\mathsf{H}}\mathbf{w}_{j}|^{2}+2\mathrm{Re}(b_{k}\mathbf{h}_{k}^{\mathsf{H}}\mathbf{w}_{k})\\
 & \hspace{-0cm}+\mathsf{R}_{k}(\{\underline{\mathbf{w}}_{i}\})-\mathsf{SINR}_{k}(\{\underline{\mathbf{w}}_{i}\})-a_{k}\sigma_{k}^{2}\Bigr),
\end{aligned}
\label{eq:f_bar MISO}
\end{equation}
where $a_{k}=\frac{\mathsf{SINR}_{k}(\{\underline{\mathbf{w}}_{i}\})}{\sum_{j=1}^{K}|\mathbf{h}_{k}^{\mathsf{H}}\underline{\mathbf{w}}_{j}|^{2}+\sigma_{k}^{2}}$
and $b_{k}=\frac{\mathsf{SINR}_{k}(\{\underline{\mathbf{w}}_{i}\})}{\mathbf{h}_{k}^{\mathsf{H}}\underline{\mathbf{w}}_{k}}$.
Disregarding constant terms in the surrogate function, we can solve
Problem \eqref{0 WSR Maximization} by iteratively solving the following
problem: 
\begin{equation}
\begin{aligned} & \underset{\{\mathbf{w}_{i}\}\in\mathcal{W}}{\text{maximize}} &  & \sum_{k=1}^{K}\omega_{k}\Bigl(-a_{k}\sum_{j=1}^{K}\bigl|\mathbf{h}_{k}^{\mathsf{H}}\mathbf{w}_{j}\bigr|^{2}+2\mathrm{Re}\left(b_{k}\mathbf{h}_{k}^{\mathsf{H}}\mathbf{w}_{k}\right)\Bigr).\end{aligned}
\label{eq:MISO-MM}
\end{equation}
This approach not only simplifies the original problem but also ensures
that each step is tractable via convex optimization. Using the Lagrangian
multipliers method, we can obtain the optimal solution to \eqref{eq:MISO-MM}
via \eqref{eq:bisection mu MISO}. We give the update step in WSR-MM
in the following with $\mu$ determined by \eqref{eq:bisection mu MISO}.

\noindent %
\noindent\fbox{\begin{minipage}[t]{1\columnwidth - 2\fboxsep - 2\fboxrule}%
\begin{center}
\labelText{\textbf{A3}}{label:MISO-MM}: The WSR-MM algorithm for
MISO beamforming \cite{zhang2021weighted}
\[
\mathbf{w}_{k}=\Bigl(\sum_{j=1}^{K}\omega_{j}a_{j}\mathbf{h}_{j}\mathbf{h}_{j}^{\mathsf{H}}+\mu\mathbf{I}\Bigr)^{\dagger}\omega_{k}b_{k}^{*}\mathbf{h}_{k}.
\]
\par\end{center}%
\end{minipage}}

\section{WSR Maximization Over MIMO Interference Channel \label{Section: MIMO}}

\subsection{System Model and Problem Formulation}

We consider a device-to-device MIMO communication system with $K$
disjoint links.\footnote{To streamline the discussion of MIMO system design alongside the MISO
scenario, certain notations will be abused in this section, though
this involves a slight repurposing of their original meanings.} We assume, in the $k$-th ($k=1,\ldots,K$) link, the transmitter
$k$ is equipped with $M_{k}^{\mathsf{t}}$ transmit antennas and
the receiver $k$ has $M_{k}^{\mathsf{r}}$ receive antennas. Denote
by $\mathbf{H}_{i,j}\in\mathbb{C}^{M_{i}^{\mathsf{r}}\times M_{j}^{\mathsf{t}}}$
the channel between receiver $i$ and transmitter $j$. Define $\mathbf{W}_{k}\in\mathbb{C}^{M_{k}^{\mathsf{t}}\times M_{k}^{\mathsf{s}}}$
($M_{k}^{\mathsf{s}}$ is the number of parallel data streams transmitted
on the $k$-th link with $M_{k}^{\mathsf{s}}\leq\min\{M_{k}^{\mathsf{t}},M_{k}^{\mathsf{r}}\}$)
as the beamforming matrix at the transmitter $k$, $\mathbf{s}_{k}\in\mathbb{C}^{M_{k}^{\mathsf{s}}}$
as the corresponding symbol vector, and $\mathbf{n}_{k}\sim\mathcal{CN}(\mathbf{0},\sigma_{k}^{2}\mathbf{I})$
as the noise. The received signal $\mathbf{y}_{k}$ at receiver $k$
is expressed as
\begin{equation}
\mathbf{y}_{k}=\mathbf{H}_{k,k}\mathbf{W}_{k}\mathbf{s}_{k}+\sum_{j=1,j\neq k}^{K}\mathbf{H}_{k,j}\mathbf{W}_{j}\mathbf{s}_{j}+\mathbf{n}_{k}.
\end{equation}
Accordingly, the data rate at the $k$-th receiver is defined by
\begin{equation}
\mathsf{R}_{k}=\log\det\left(\mathbf{I}+\mathbf{W}_{k}^{\mathsf{H}}\mathbf{H}_{k,k}^{\mathsf{H}}\mathbf{F}_{k}^{-1}\mathbf{H}_{k,k}\mathbf{W}_{k}\right)
\end{equation}
where the interference-plus-noise matrix is denoted as
\begin{equation}
\mathbf{F}_{k}=\sum_{j=1,j\neq k}^{K}\mathbf{H}_{k,j}\mathbf{W}_{j}\mathbf{W}_{j}^{\mathsf{H}}\mathbf{H}_{k,j}^{\mathsf{H}}+\sigma_{k}^{2}\mathbf{I}.
\end{equation}
Our goal is to maximize the WSR of the system by designing the beamformer
matrices $\{\mathbf{W}_{i}\}$, thereby solving the following maximization
problem:
\begin{equation}
\begin{aligned} & \underset{\{\mathbf{W}_{i}\}\in\mathcal{W}}{\text{maximize}} &  & f\left(\{\mathbf{W}_{i}\}\right)=\sum_{k=1}^{K}\omega_{k}\mathsf{R}_{k},\end{aligned}
\label{MIMO: WSR Maximization Problem}
\end{equation}
where the transmit power limit constraint is given by
\begin{equation}
\mathcal{W}=\left\{ \{\mathbf{W}_{i}\}\mid\left\Vert \mathbf{W}_{i}\right\Vert _{\mathsf{}}^{2}\leq P_{i},\ i=1,\ldots,K\right\} 
\end{equation}
with $P_{i}$ representing the power budget at the $i$-th transmitter.

In the following, we will introduce the WMMSE \cite{shi2011iteratively},
WSR-FP \cite{shen2019optimization}, and WSR-MM \cite{zhang2021weighted}
methods for MIMO beamforming.

\subsection{WMMSE Algorithms\label{Section: WMMSE for WSR Maximization MIMO}}

We consider the WMMSE approach proposed in \cite{shi2011iteratively}.
Let $\mathbf{L}_{k}\in\mathbb{C}^{M_{k}^{\mathsf{r}}\times M_{k}^{\mathsf{s}}}$
be the receive beamformer of receiver $k$, $k=1,\ldots,K$. Under
the independence assumption of $\mathbf{s}_{k}$'s and $\mathbf{n}_{k}$'s,
the MSE matrix between the estimated signal and the original signal
at receiver $k$ is given by 
\begin{equation}
\begin{aligned}\mathbf{E}_{k}=\, & \mathbb{E}\biggl[\Bigl(\mathbf{L}_{k}^{\mathsf{H}}\bigl(\mathbf{H}_{k,k}\mathbf{W}_{k}\mathbf{s}_{k}+\sum_{j=1,j\neq k}^{K}\mathbf{H}_{k,j}\mathbf{W}_{j}\mathbf{s}_{j}+\mathbf{n}_{k}\bigr)-\mathbf{s}_{k}\Bigr)\cdot\\
 & \Bigl(\mathbf{L}_{k}^{\mathsf{H}}\bigl(\mathbf{H}_{k,k}\mathbf{W}_{k}\mathbf{s}_{k}+\sum_{j=1,j\neq k}^{K}\mathbf{H}_{k,j}\mathbf{W}_{j}\mathbf{s}_{j}+\mathbf{n}_{k}\bigr)-\mathbf{s}_{k}\Bigr)^{\mathsf{H}}\biggr]\\
=\, & \left(\mathbf{L}_{k}^{\mathsf{H}}\mathbf{H}_{k,k}\mathbf{W}_{k}-\mathbf{I}\right)\left(\mathbf{L}_{k}^{\mathsf{H}}\mathbf{H}_{k,k}\mathbf{W}_{k}-\mathbf{I}\right)^{\mathsf{H}}\\
 & +\sum_{j=1,j\neq k}^{K}\mathbf{L}_{k}^{\mathsf{H}}\mathbf{H}_{k,j}\mathbf{W}_{j}\mathbf{W}_{j}^{\mathsf{H}}\mathbf{H}_{k,j}^{\mathsf{H}}\mathbf{L}_{k}+\sigma_{k}^{2}\mathbf{L}_{k}^{\mathsf{H}}\mathbf{L}_{k}.
\end{aligned}
\end{equation}

\begin{prop}[\cite{shi2011iteratively}]
\label{thm: WMMSE MIMO} Let $\mathbf{M}_{1},\ldots,\mathbf{M}_{K}\succeq\mathbf{0}$
be the weight matrices. The WSR maximization problem \eqref{MIMO: WSR Maximization Problem}
is equivalent to the following weighted sum-MSE minimization problem
\begin{equation}
\begin{aligned} & \underset{\{\mathbf{L}_{i}\},\{\mathbf{M}_{i}\},\{\mathbf{W}_{i}\}\in\mathcal{W}}{\text{maximize}} &  & \sum_{k=1}^{K}\omega_{k}\left(\log\det\left(\mathbf{M}_{k}\right)-\mathrm{tr}\left(\mathbf{M}_{k}\mathbf{E}_{k}\right)\right),\end{aligned}
\label{MIMO WSMSE}
\end{equation}
in the sense that they attain the identical optimal solution. 
\end{prop}
Similar to the MISO case, based on BCA, update rules of the WMMSE
algorithm for Problem \eqref{MIMO WSMSE} are summarized below.

\noindent %
\noindent\fbox{\begin{minipage}[t]{1\columnwidth - 2\fboxsep - 2\fboxrule}%
\begin{center}
\labelText{\textbf{A4}}{label:MIMO-WMMSE}: A WMMSE algorithm for
MIMO beamforming \cite{shi2011iteratively}
\begin{align*}
\text{S1: } & \mathbf{L}_{k}=(\underline{\mathbf{F}}_{k}+\mathbf{H}_{k,k}\underline{\mathbf{W}}_{k}\underline{\mathbf{W}}_{k}^{\mathsf{H}}\mathbf{H}_{k,k}^{\mathsf{H}})^{-1}\mathbf{H}_{k,k}\underline{\mathbf{W}}_{k}\\
\text{S2: } & \mathbf{M}_{k}=\bigl(\sum_{j=1}^{K}\underline{\mathbf{L}}_{k}^{\mathsf{H}}\mathbf{H}_{k,j}\underline{\mathbf{W}}_{j}\underline{\mathbf{W}}_{j}^{\mathsf{H}}\mathbf{H}_{k,j}^{\mathsf{H}}\underline{\mathbf{L}}_{k}-\underline{\mathbf{L}}_{k}^{\mathsf{H}}\mathbf{H}_{k,k}\underline{\mathbf{W}}_{k}\\
 & \hspace{3.5cm}-\underline{\mathbf{W}}_{k}^{\mathsf{H}}\mathbf{H}_{k,k}^{\mathsf{H}}\underline{\mathbf{L}}_{k}+\sigma_{k}^{2}\underline{\mathbf{L}}_{k}^{\mathsf{H}}\underline{\mathbf{L}}_{k}+\mathbf{I}\bigr)^{-1}\\
\text{S3: } & \mathbf{W}_{k}=\bigl(\sum_{j=1}^{K}\omega_{j}\mathbf{H}_{k,j}^{\mathsf{H}}\underline{\mathbf{L}}_{j}\underline{\mathbf{M}}_{j}\underline{\mathbf{L}}_{j}^{\mathsf{H}}\mathbf{H}_{k,j}+\mu_{k}\mathbf{I}\bigr)^{\dagger}\omega_{k}\mathbf{H}_{k,k}^{\mathsf{H}}\underline{\mathbf{L}}_{k}\underline{\mathbf{M}}_{k}.
\end{align*}
\par\end{center}%
\end{minipage}}

In \nameref{label:MIMO-WMMSE}, $\mu_{k}$ for $k=1,\ldots,K$ is
optimally determined by
\begin{equation}
\mu_{k}=\min\bigl\{\mu_{k}\geq0:\left\Vert \mathbf{W}_{k}(\mu_{k})\right\Vert _{\mathsf{}}^{2}\leq P_{k}\bigr\}.\label{eq:bisection mu MIMO}
\end{equation}

\subsection{WSR-FP Algorithms\label{Section: FP for WSR Maximization MIMO}}

We discuss the WSR-FP approach proposed in \cite{shen2019optimization}.
To begin with, we first present the matrix Lagrangian dual transform
and the matrix quadratic transform. 
\begin{prop}[Matrix Lagrangian dual transform \cite{shen2019optimization}]
\label{FP: Matrix Lagrangian dual transform} Given matrix-ratios
$\sqrt{\mathbf{C}}_{k}^{\mathsf{H}}(\mathbf{X})\mathbf{D}_{k}^{-1}(\mathbf{X})\sqrt{\mathbf{C}}_{k}(\mathbf{X})$
with $\mathbf{C}_{k}(\mathbf{X})\succeq\mathbf{0}$ and $\mathbf{D}_{k}(\mathbf{X})\succ\mathbf{0}$
for $k=1,\ldots,K$, the following problem: 
\begin{equation}
\begin{aligned} & \underset{\mathbf{X}\in\mathcal{X}}{\text{maximize}} &  & \sum_{k=1}^{K}\omega_{k}\log\det\bigl(\mathbf{I}+\sqrt{\mathbf{C}}_{k}^{\mathsf{H}}(\mathbf{X})\mathbf{D}_{k}^{-1}(\mathbf{X})\sqrt{\mathbf{C}}_{k}(\mathbf{X})\bigr),\end{aligned}
\label{Matrix FP: sum-of-log}
\end{equation}
is equivalent to 
\begin{equation}
\begin{aligned} & \negthickspace\negthickspace\underset{\{\boldsymbol{\Gamma}_{i}\},\mathbf{X}\in\mathcal{X}}{\text{maximize}} &  & \negthickspace\negthickspace\sum_{k=1}^{K}\omega_{k}\Bigl(\log\det\left(\mathbf{I}+\boldsymbol{\Gamma}_{k}\right)-\mathrm{tr}\left(\boldsymbol{\Gamma}_{k}\right)+\mathrm{tr}\bigl((\mathbf{I}\\
 &  &  & \hspace{-1.3cm}+\boldsymbol{\Gamma}_{k})\sqrt{\mathbf{C}}_{k}^{\mathsf{H}}(\mathbf{X})\bigl(\mathbf{C}_{k}(\mathbf{X})+\mathbf{D}_{k}(\mathbf{X})\bigr)^{-1}\sqrt{\mathbf{C}}_{k}(\mathbf{X})\bigr)\Bigr),
\end{aligned}
\negthickspace\label{Matrix FP: Lagrangian transformed}
\end{equation}
in the sense that they attain the identical optimal solution with
the identical optimal objective value. 
\end{prop}
\begin{prop}[Matrix quadratic transform \cite{shen2019optimization}]
\label{FP: Matrix quadratic transform} Given nondecreasing matrix
functions $q_{k}\left(\cdot\right)$ such that $q_{k}\left(\mathbf{Z}_{1}\right)\geq q_{k}\left(\mathbf{Z}_{2}\right)$
if $\mathbf{Z}_{1}\succeq\mathbf{Z}_{2}$ and ratios $\sqrt{\mathbf{C}}_{k}^{\mathsf{H}}(\mathbf{X})\mathbf{D}_{k}^{-1}(\mathbf{X})\sqrt{\mathbf{C}}_{k}(\mathbf{X})$
with $\mathbf{C}_{k}(\mathbf{X})\succeq\mathbf{0}$ and $\mathbf{D}_{k}(\mathbf{X})\succ\mathbf{0}$
for $k=1,\ldots,K$, the following problem: 
\begin{equation}
\begin{aligned} & \underset{\mathbf{X}\in\mathcal{X}}{\text{maximize}} &  & \sum_{k=1}^{K}q_{k}\bigl(\sqrt{\mathbf{C}}_{k}^{\mathsf{H}}(\mathbf{X})\mathbf{D}_{k}^{-1}(\mathbf{X})\sqrt{\mathbf{C}}_{k}(\mathbf{X})\bigr)\end{aligned}
,\label{Matrix FP: sum-of-functions-of-matrix-ratio}
\end{equation}
is equivalent to 
\begin{equation}
\begin{aligned} & \underset{\{\boldsymbol{\Phi}_{i}\},\mathbf{X}\in\mathcal{X}}{\text{maximize}} &  & \sum_{k=1}^{K}q_{k}\Bigl(2\mathrm{Re}\bigl(\sqrt{\mathbf{C}}_{k}^{\mathsf{H}}(\mathbf{X})\boldsymbol{\Phi}_{k}\bigr)-\boldsymbol{\Phi}_{k}^{\mathsf{H}}\mathbf{D}_{k}(\mathbf{X})\boldsymbol{\Phi}_{k}\Bigr),\end{aligned}
\label{Matrix FP: quadratic transformed}
\end{equation}
in the sense that they attain the identical optimal solution with
the identical optimal objective value. 
\end{prop}
By applying Proposition \ref{FP: Matrix Lagrangian dual transform}
to Problem \eqref{MIMO: WSR Maximization Problem}, and considering
$\sqrt{\mathbf{C}}_{k}(\mathbf{X})=\mathbf{W}_{k}^{\mathsf{H}}\mathbf{H}_{k,k}^{\mathsf{H}}$
and $\mathbf{D}_{k}(\mathbf{X})=\sum_{j=1,j\neq k}^{K}\mathbf{H}_{k,j}\mathbf{W}_{j}\mathbf{W}_{j}^{\mathsf{H}}\mathbf{H}_{k,j}^{\mathsf{H}}+\sigma_{k}^{2}\mathbf{I}$,
we get the following problem with auxiliary variables $\{\boldsymbol{\Gamma}_{i}\}$:
\begin{equation}
\begin{aligned} & \underset{\{\boldsymbol{\Gamma}_{i}\},\{\mathbf{W}_{i}\}\in\mathcal{W}}{\text{maximize}} &  & \hspace{-0.2cm}\sum_{k=1}^{K}\omega_{k}\Bigl(\log\det(\mathbf{I}+\boldsymbol{\Gamma}_{k})-\mathrm{tr}(\boldsymbol{\Gamma}_{k})+\mathrm{tr}\bigl((\mathbf{I}\\
 &  &  & \hspace{-2.5cm}+\boldsymbol{\Gamma}_{k})\mathbf{H}_{k,k}\mathbf{W}_{k}\bigl(\sum_{j=1}^{K}\mathbf{H}_{k,j}\mathbf{W}_{j}\mathbf{W}_{j}^{\mathsf{H}}\mathbf{H}_{k,j}^{\mathsf{H}}+\sigma_{k}^{2}\mathbf{I}\bigr)^{-1}\mathbf{W}_{k}^{\mathsf{H}}\mathbf{H}_{k,k}^{\mathsf{H}}\bigr)\Bigr),
\end{aligned}
\label{FP: f_LD MIMO}
\end{equation}
The ratios in Problem \eqref{FP: f_LD MIMO} can be further decoupled
by applying Proposition \ref{FP: Matrix quadratic transform} with
$q_{k}(\mathbf{Z})=\mathrm{tr}\left(\left(\mathbf{I}+\boldsymbol{\Gamma}_{k}\right)\mathbf{Z}\right)$,
$\sqrt{\mathbf{C}}_{k}(\mathbf{x})=\sqrt{\omega_{k}}\mathbf{W}_{k}^{\mathsf{H}}\mathbf{H}_{k,k}^{\mathsf{H}}$,
and $\mathbf{D}_{k}(\mathbf{x})=\sum_{j=1,j\neq k}^{K}\mathbf{H}_{k,j}\mathbf{W}_{j}\mathbf{W}_{j}^{\mathsf{H}}\mathbf{H}_{k,j}^{\mathsf{H}}+\sigma_{k}^{2}\mathbf{I}$,
leading to an equivalent problem with further introduced auxiliary
variables $\{\boldsymbol{\Phi}_{i}\}$:
\begin{equation}
\begin{aligned} & \underset{\{\boldsymbol{\Gamma}_{i}\},\{\boldsymbol{\Phi}_{i}\},\{\mathbf{W}_{i}\}\in\mathcal{W}}{\text{maximize}} &  & \hspace{-0.2cm}\sum_{k=1}^{K}\omega_{k}\biggl(\log\det(\mathbf{I}+\boldsymbol{\Gamma}_{k})\\
 &  &  & \hspace{-0.4cm}\hspace{-0.5cm}\hspace{-0.45cm}\hspace{-0.5cm}\hspace{-0.45cm}-\mathrm{tr}(\boldsymbol{\Gamma}_{k})+\mathrm{tr}\Bigl((\mathbf{I}+\boldsymbol{\Gamma}_{k})\bigl(2\mathrm{Re}(\sqrt{\omega_{k}}\mathbf{H}_{k,k}\mathbf{W}_{k}\boldsymbol{\Phi}_{k})\\
 &  &  & \hspace{-0.4cm}\hspace{-0.5cm}\hspace{-0.45cm}\hspace{-0.3cm}-\boldsymbol{\Phi}_{k}^{\mathsf{H}}\bigl(\sum_{j=1}^{K}\mathbf{H}_{k,j}\mathbf{W}_{j}\mathbf{W}_{j}^{\mathsf{H}}\mathbf{H}_{k,j}^{\mathsf{H}}+\sigma_{k}^{2}\mathbf{I}\bigr)\boldsymbol{\Phi}_{k}\bigr)\Bigr)\biggr).
\end{aligned}
\label{MIMO FP: Final Problem}
\end{equation}
Then based on BCA, variable updates in a WSR-FP algorithm are summarized
in the following ($\mu_{k}$ is determined by \eqref{eq:bisection mu MIMO}).

\noindent %
\noindent\fbox{\begin{minipage}[t]{1\columnwidth - 2\fboxsep - 2\fboxrule}%
\begin{center}
\labelText{\textbf{A5}}{label:MIMO-FP}: A WSR-FP algorithm for
MIMO beamforming \cite{shen2019optimization}
\begin{align*}
\text{S1: } & \boldsymbol{\Phi}_{k}=\bigl(\underline{\mathbf{F}}_{k}+\mathbf{H}_{k,k}\underline{\mathbf{W}}_{k}\underline{\mathbf{W}}_{k}^{\mathsf{H}}\mathbf{H}_{k,k}^{\mathsf{H}}\bigr)^{-1}\sqrt{\omega_{k}}\mathbf{H}_{k,k}\underline{\mathbf{W}}_{k},\\
\text{S2: } & \boldsymbol{\Gamma}_{k}=\underline{\mathbf{W}}_{k}^{\mathsf{H}}\mathbf{H}_{k,k}^{\mathsf{H}}\underline{\mathbf{F}}_{k}^{-1}\mathbf{H}_{k,k}\underline{\mathbf{W}}_{k},\\
\text{S3: } & \mathbf{W}_{k}=\Bigl(\sum_{j=1}^{K}\mathbf{H}_{k,j}^{\mathsf{H}}\underline{\boldsymbol{\Phi}}_{j}(\mathbf{I}+\underline{\boldsymbol{\Gamma}}_{j})\underline{\boldsymbol{\Phi}}_{j}^{\mathsf{H}}\mathbf{H}_{k,j}+\mu_{k}\mathbf{I}\Bigr)^{\dagger}\cdot\\
 & \hspace{1cm}\sqrt{\omega_{k}}\mathbf{H}_{k,k}^{\mathsf{H}}\underline{\boldsymbol{\Phi}}_{k}(\mathbf{I}+\underline{\boldsymbol{\Gamma}}_{k}).
\end{align*}
\par\end{center}%
\end{minipage}}

\subsection{WSR-MM Algorithms \label{Section: MM for WSR Maximization MIMO}}

In this section, we introduce the WSR-MM approach \cite{zhang2021weighted}
for MIMO beamforming.
\begin{prop}[\cite{zhang2021weighted}]
\label{Prop: Matrix MM} Given $\log\det\left(\mathbf{I}+\mathbf{X}^{\mathsf{H}}\mathbf{Z}^{-1}\mathbf{X}\right)$
with $\mathbf{X}\in\mathbb{C}^{M\times N}$ and $\mathbf{Z}\succ\mathbf{0}$,
at $\left(\underline{\mathbf{X}},\underline{\mathbf{Z}}\right)$ we
have
\begin{equation}
\begin{aligned}\log\det(\mathbf{I}+\mathbf{X}^{\mathsf{H}}\mathbf{Z}^{-1}\mathbf{X})\geq & \log\det(\mathbf{I}+\underline{\mathbf{X}}^{\mathsf{H}}\underline{\mathbf{Z}}^{-1}\underline{\mathbf{X}})\\
 & \hspace{-2.5cm}-{\rm tr}(\underline{\mathbf{X}}^{\mathsf{H}}\underline{\mathbf{Z}}^{-1}\underline{\mathbf{X}})+2{\rm {\rm Re}}\bigl({\rm tr}\bigl(\underline{\mathbf{X}}^{\mathsf{H}}\underline{\mathbf{Z}}^{-1}\mathbf{X}\bigr)\bigr)\\
 & \hspace{-2.5cm}-\mathrm{tr}\bigl((\underline{\mathbf{Y}}+\underline{\mathbf{X}}\underline{\mathbf{X}}^{\mathsf{H}})^{-1}\underline{\mathbf{X}}\underline{\mathbf{X}}^{\mathsf{H}}\underline{\mathbf{Z}}^{-1}(\mathbf{Z}+\mathbf{X}\mathbf{X}^{\mathsf{H}})\bigr),
\end{aligned}
\end{equation}
where the equality is attained when $\left(\mathbf{X},\mathbf{Z}\right)=\left(\underline{\mathbf{X}},\underline{\mathbf{Z}}\right)$. 
\end{prop}
By applying Proposition \ref{Prop: Matrix MM} to $f\left(\{\mathbf{W}_{i}\}\right)$
with $\mathbf{X}=\mathbf{H}_{k,k}\mathbf{W}_{k}$ and $\mathbf{Z}=\mathbf{F}_{k}$,
we construct a surrogate function at $\{\underline{\mathbf{W}_{i}}\}$
as follows:
\begin{equation}
\begin{aligned} & \ell\left(\{\mathbf{W}_{i}\},\{\underline{\mathbf{W}_{i}}\}\right)\\
=\: & \sum_{k=1}^{K}\omega_{k}\Bigl(-\mathrm{tr}\Bigl(\sum_{j=1}^{K}\mathbf{W}_{j}^{\mathsf{H}}\mathbf{H}_{k,j}^{\mathsf{H}}\mathbf{A}_{k}\mathbf{H}_{k,j}\mathbf{W}_{j}\Bigr)\\
 & +2{\rm {\rm Re}}\bigl({\rm tr}\bigl(\mathbf{B}_{k}\mathbf{H}_{k,k}\mathbf{W}_{k}\bigr)\bigr)\\
 & +\log\det(\mathbf{I}+\underline{\mathbf{W}}_{k}^{\mathsf{H}}\mathbf{H}_{k,k}^{\mathsf{H}}\underline{\mathbf{F}}_{k}^{-1}\mathbf{H}_{k,k}\underline{\mathbf{W}}_{k})\\
 & -{\rm tr}(\underline{\mathbf{W}}_{k}^{\mathsf{H}}\mathbf{H}_{k,k}^{\mathsf{H}}\underline{\mathbf{F}}_{k}^{-1}\mathbf{H}_{k,k}\underline{\mathbf{W}}_{k})-\mathrm{tr}\Bigl(\sigma_{k}^{2}\mathbf{A}_{k}\Bigr)\Bigr),
\end{aligned}
\label{eq: f_bar MIMO}
\end{equation}
where $\mathbf{A}_{k}=(\underline{\mathbf{F}}_{k}+\mathbf{H}_{k,k}\underline{\mathbf{W}}_{k}\underline{\mathbf{W}}_{k}^{\mathsf{H}}\mathbf{H}_{k,k}^{\mathsf{H}})^{-1}\mathbf{H}_{k,k}\underline{\mathbf{W}}_{k}\mathbf{B}_{k}$
and $\mathbf{B}_{k}=\underline{\mathbf{W}}_{k}^{\mathsf{H}}\mathbf{H}_{k,k}^{\mathsf{H}}\underline{\mathbf{F}}_{k}^{-1}$.
Based on the principal of MM, we solve Problem \eqref{MIMO: WSR Maximization Problem}
by successively solving the following problem: 
\begin{equation}
\begin{aligned} & \underset{\{\mathbf{W}_{i}\}\in\mathcal{W}}{\text{maximize}} &  & \sum_{k=1}^{K}\Bigl(-\mathrm{tr}\bigl(\mathbf{W}_{k}^{\mathsf{H}}\sum_{j=1}^{K}\omega_{j}\mathbf{H}_{j,k}^{\mathsf{H}}\mathbf{A}_{j}\mathbf{H}_{j,k}\mathbf{W}_{k}\bigr)\\
 &  &  & \hspace{2cm}+2\omega_{k}\mathrm{tr}\left(\mathbf{B}_{k}\mathbf{H}_{k,k}\mathbf{W}_{k}\right)\Bigr),
\end{aligned}
\end{equation}
whose solution can be obtained via the Lagrangian multipliers method
\eqref{eq:bisection mu MIMO}. WSR-MM is given in \nameref{label:MIMO-MM}
with $\mu_{k}$ defined in \eqref{eq:bisection mu MIMO}.

\noindent %
\noindent\fbox{\begin{minipage}[t]{1\columnwidth - 2\fboxsep - 2\fboxrule}%
\begin{center}
\labelText{\textbf{A6}}{label:MIMO-MM}: The WSR-MM algorithm for
MIMO beamforming \cite{zhang2021weighted}
\[
\mathbf{W}_{k}=\Bigl(\sum_{j=1}^{K}\omega_{j}\mathbf{H}_{j,k}^{\mathsf{H}}\mathbf{A}_{j}\mathbf{H}_{j,k}+\mu_{k}\mathbf{I}\Bigr)^{\dagger}\omega_{k}\mathbf{H}_{k,k}^{\mathsf{H}}\mathbf{B}_{k}^{\mathsf{H}}.
\]
\par\end{center}%
\end{minipage}}

\section{Connections Among WMMSE, WSR-FP, and WSR-MM \label{Section: connection} }

The connection between WMMSE and WSR-FP has been elucidated in \cite{shen2019optimization}.
Given that the weighted sum-MSE minimization problem \eqref{MIMO WSMSE}
emerges as a particular instance of applying Proposition \ref{FP: Matrix quadratic transform}
to Problem \eqref{FP: f_LD MIMO}, it is evident that WMMSE algorithms
represent specific cases within the broader scope of WSR-FP algorithms.
Our subsequent discussion mainly focus on the connections between
WSR-FP/WMMSE with WSR-MM algorithms. Our analysis will be centered
on the MIMO case, though it should be noted that the findings are
equally applicable to the MISO scenario. We will initiate by establishing
a general result that delineates the process of converting a three-block
BCA into an MM algorithm, laying the groundwork for the subsequent
analysis.  

\subsection{Mapping A Specific BCA Algorithm to An MM Algorithm }

Consider the optimization variables $u$, $v$, and $x$ in Problem
\eqref{eq:BCA}, which are updated cyclically and non-repetitively
according to \eqref{eq:BCA update}. It is postulated that $\max_{u\in\mathcal{U},v\in\mathcal{V}}g(u,v,x)=f(x)$,
with the subsequent assumption that the update for $u^{(t+1)}$ is
solely dependent on $x^{(t)}$ and is independent of $v^{(t)}$, and
$v^{(t+1)}$ is updated based on both $x^{(t)}$ and $u^{(t+1)}$.
Under these premises, the triple update steps in \eqref{eq:BCA update}
can be reformulated into a dual-step process:
\begin{equation}
\begin{cases}
\left(u^{(t+1)},v^{(t+1)}\right)\in\arg\max_{u\in\mathcal{U},v\in\mathcal{V}}g(u,v,x^{(t)})\\
x^{(t+1)}\in\arg\max_{x\in\mathcal{X}}g(u^{(t+1)},v^{(t+1)},x).
\end{cases}\label{eq:BCA-1}
\end{equation}
This reformulation ensures that for all $(x,x^{(t)})\in{\cal X}$,
the following relations hold true:
\begin{align*}
g(u^{(t+1)},v^{(t+1)},x)\leq\max_{u\in\mathcal{U},v\in\mathcal{V}}g(u,v,x)=f(x),
\end{align*}
and
\[
g(u^{(t+1)},v^{(t+1)},x^{(t)})=\max_{u\in\mathcal{U},v\in\mathcal{V}}g(u,v,x^{(t)})=f(x^{(t)}),
\]
from which we deduce that $g(u^{(t+1)},v^{(t+1)},x)$ serves as a
surrogate function for $f(x)$ at iterate $x^{(t)}$. Consequently,
\eqref{eq:BCA-1} may be interpreted as an MM algorithm step. It is
noteworthy that \eqref{eq:BCA-1} can be further distilled into a
fixed-point iteration: $x^{(t+1)}\in\arg\max_{x\in\mathcal{X}}g(\arg\max_{u\in\mathcal{U},v\in\mathcal{V}}g(u,v,x^{(t)}),x)$.

\subsection{Connections Between WMMSE/WSR-FP and WSR-MM}

Recall that both the WMMSE and WSR-FP are three-block BCA algorithms.
The result given in last section facilitates the comprehension of
their relationship with the WSR-MM approach. Considering the WMMSE
as formulated in \nameref{label:MIMO-WMMSE}, and incorporating $\mathbf{L}_{k}$
into $\mathbf{M}_{k}$, yields the following expression:
\begin{equation}
\begin{aligned}\mathbf{M}_{k} & =\bigl(\mathbf{I}-\underline{\mathbf{W}}_{k}^{\mathsf{H}}\mathbf{H}_{k,k}^{\mathsf{H}}(\underline{\mathbf{F}}_{k}+\mathbf{H}_{k,k}\underline{\mathbf{W}}_{k}\underline{\mathbf{W}}_{k}^{\mathsf{H}}\mathbf{H}_{k,k}^{\mathsf{H}})^{-1}\mathbf{H}_{k,k}\underline{\mathbf{W}}_{k}\bigr)^{-1}\\
 & =\mathbf{I}+\underline{\mathbf{W}}_{k}^{\mathsf{H}}\mathbf{H}_{k,k}^{\mathsf{H}}\underline{\mathbf{F}}_{k}^{-1}\mathbf{H}_{k,k}\underline{\mathbf{W}}_{k},
\end{aligned}
\end{equation}
where the second line is due to the Woodbury matrix identity \cite{higham2002accuracy}.
Define $\mathbf{A}_{k}=\mathbf{L}_{k}\mathbf{M}_{k}\mathbf{L}_{k}^{\mathsf{H}}$
and $\mathbf{B}_{k}=\mathbf{M}_{k}^{\mathsf{H}}\mathbf{L}_{k}^{\mathsf{H}}$.
We can recover \nameref{label:MIMO-MM} from \nameref{label:MIMO-WMMSE},
indicating that a WMMSE algorithm can equate to the WSR-MM. However,
the alternative formulation of WMMSE presented in \eqref{eq:WMMSE another update order}
defies this equivalence, as it does not conform to the update rule
in \eqref{eq:BCA-1}. The connection between WSR-FP and WSR-MM mirrors
that of WMMSE and WSR-MM. As demonstrated in \cite{shen2019optimization},
the update procedures in WSR-FP in \nameref{label:MIMO-FP} are profoundly
linked to a surrogate function. Advancing this concept, we show that
the specialized WSR-FP algorithm \nameref{label:MIMO-FP} corresponds
explicitly to the WSR-MM algorithm \nameref{label:MIMO-MM}. By substituting
$\{\boldsymbol{\Gamma}_{i}\}$ and $\{\boldsymbol{\Phi}_{i}\}$ into
the update step of $\mathbf{W}_{k}$ in \nameref{label:MIMO-FP} and
defining $\mathbf{A}_{k}=\frac{1}{\omega_{k}}\boldsymbol{\Phi}_{k}(\mathbf{I}+\boldsymbol{\Gamma}_{k})\boldsymbol{\Phi}_{k}^{\mathsf{H}}$
and $\mathbf{B}_{k}=\frac{1}{\sqrt{\omega_{k}}}(\mathbf{I}+\boldsymbol{\Gamma}_{k}^{\mathsf{H}})\boldsymbol{\Phi}_{k}^{\mathsf{H}}$,
we obtain \nameref{label:MIMO-MM}. It is important to note, however,
that certain variants of WSR-FP do not align with the MM framework.
For instance, replacing \textbf{S1} in \nameref{label:MISO-FP} with
\eqref{eq:conventional gamma}, which entails a conventional BCA update
rule, results in an algorithm that diverges from the MM approach.
The relationships among WMMSE, WSR-FP, and WSR-MM are summarized in
Fig. \ref{fig:Connections-among-WMMSE,}. 

\begin{figure}[t]
\begin{centering}
\includegraphics[width=0.6\columnwidth]{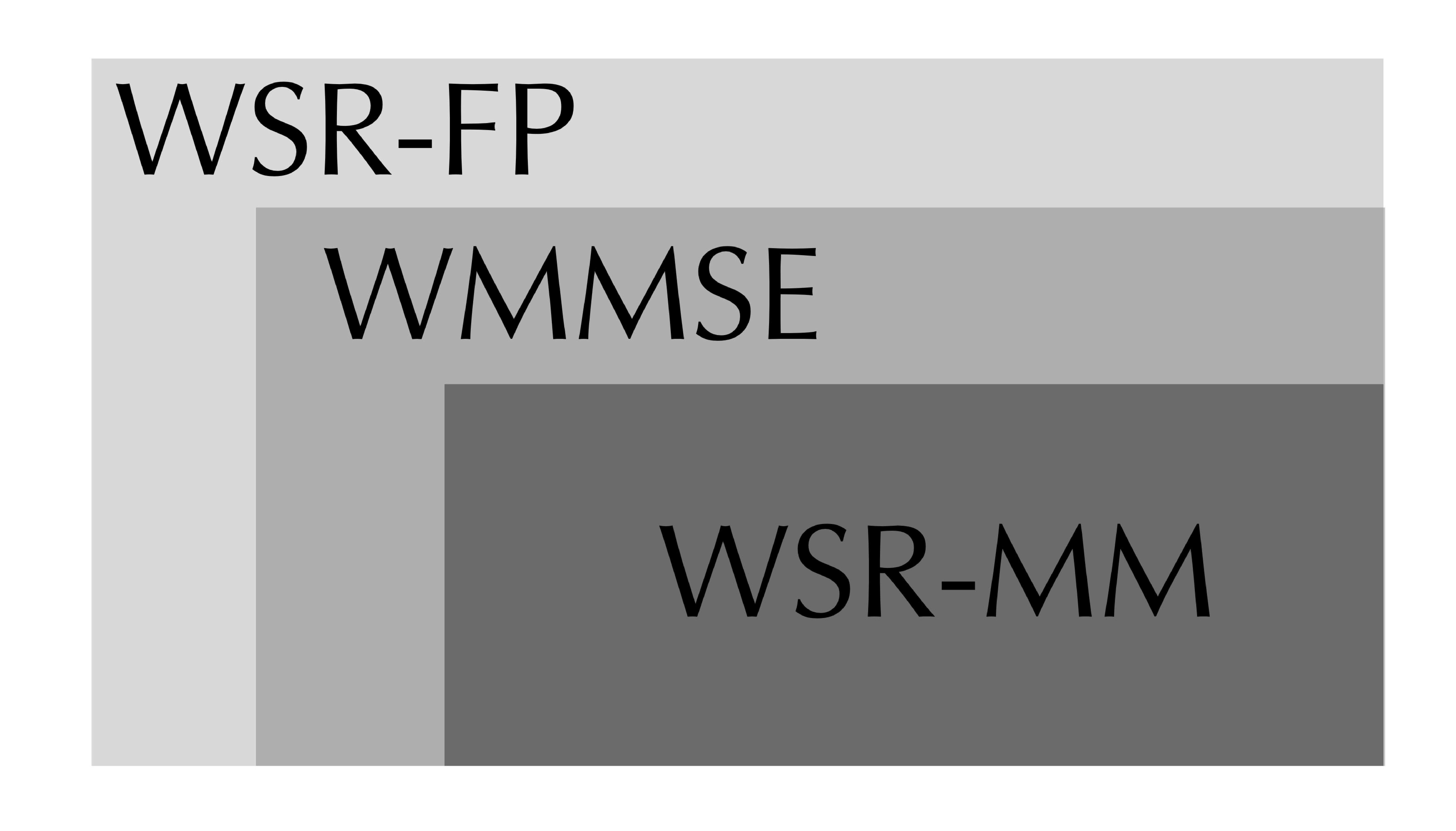}
\par\end{centering}
\caption{Relations among WMMSE, WSR-FP, and WSR-MM.\label{fig:Connections-among-WMMSE,}}
\end{figure}

\subsection{Bridging Equivalent Transforms and Surrogate Functions\label{subsec:Relating-Equivalent-Transforms}}

In this section, we elucidate that within the WSR-FP algorithm, the
refinement of auxiliary variables through both equivalent transforms---the
Lagrangian dual transform \eqref{FP: f_LD MIMO} and the quadratic
transform \eqref{MIMO FP: Final Problem}---can be interpreted as
methodologies for devising surrogate functions in the context of WSR-MM.
Prior to this, we introduce two pertinent lemmas.
\begin{lem}[\cite{zhang2021weighted}]
\label{1 Transform: MM-log-MIMO} Given $\log\det(\mathbf{Z})$ with
$\mathbf{Z}\succ\mathbf{0}$, at $\underline{\mathbf{Z}}$ we have
\begin{equation}
\log\det(\mathbf{Z})\geq\log\det(\underline{\mathbf{Z}})+{\rm tr}\bigl(\mathbf{I}-\underline{\mathbf{Z}}\mathbf{Z}^{-1}\bigr),
\end{equation}
where the equality is attained when $\mathbf{Z}=\underline{\mathbf{Z}}$. 
\end{lem}
\begin{lem}[\cite{zhang2021weighted}]
\label{1 Transform: MM-quadratic-MIMO} Given a nondecreasing matrix
function $\mathbf{Q}(\cdot)$ such that $\mathbf{Q}\left(\mathbf{Z}_{1}\right)\succeq\mathbf{Q}\left(\mathbf{Z}_{2}\right)$
if $\mathbf{Z}_{1}\succeq\mathbf{Z}_{2}$, then for $\mathbf{Q}\left(\mathbf{Z}_{1}^{\mathsf{H}}\mathbf{Z}_{2}^{-1}\mathbf{Z}_{1}\right)$
with $\mathbf{Z}_{1}\in\mathbb{C}^{n\times n}$ and $\mathbf{Z}_{2}\succ\mathbf{0}$,
at $(\underline{\mathbf{Z}_{1}},\underline{\mathbf{Z}_{2}})$ we have
\begin{equation}
\mathbf{Q}\left(\mathbf{Z}_{1}^{\mathsf{H}}\mathbf{Z}_{2}^{-1}\mathbf{Z}_{1}\right)\succeq\mathbf{Q}\left(2\mathrm{Re}\left(\underline{\mathbf{Z}_{1}^{\mathsf{H}}}\underline{\mathbf{Z}_{2}^{-1}}\mathbf{Z}_{1}\right)-\underline{\mathbf{Z}_{1}^{\mathsf{H}}}\underline{\mathbf{Z}_{2}^{-1}}\mathbf{Z}_{2}\underline{\mathbf{Z}_{2}^{-1}}\underline{\mathbf{Z}_{1}}\right),
\end{equation}
where the equality is attained when $\left(\mathbf{Z}_{1},\mathbf{Z}_{2}\right)=\left(\underline{\mathbf{Z}_{1}},\underline{\mathbf{Z}_{2}}\right)$. 
\end{lem}
As shown in \cite{zhang2021weighted}, the surrogate function $\ell$
in Section \ref{Section: MM for WSR Maximization MIMO} can be conceptualized
as being derived through two sequential operations anchored in Lemmas
\ref{1 Transform: MM-log-MIMO} and \ref{1 Transform: MM-quadratic-MIMO}.
We present the following two propositions to demonstrate how the updates
of the auxiliary variables in the WSR-FP algorithm (refer to \nameref{label:MIMO-FP})
correlate with the aforementioned lemmas.
\begin{prop}
\label{thm:Matrix Lagrangian dual MM} Updating variables $\{\boldsymbol{\Gamma}_{i}\}$
in Problem \eqref{Matrix FP: Lagrangian transformed} by BCA can be
seen as constructing a surrogate function based on Lemma \ref{1 Transform: MM-log-MIMO}
for the objective in Problem \eqref{Matrix FP: sum-of-log} by MM.
\end{prop}
\begin{IEEEproof}
With $\mathbf{X}$ fixed, $\{\boldsymbol{\Gamma}_{i}\}$ is optimally
determined by
\begin{equation}
\boldsymbol{\Gamma}_{k}=\sqrt{\mathbf{C}}_{k}^{\mathsf{H}}(\underline{\mathbf{X}})\left(\mathbf{D}_{k}(\underline{\mathbf{X}})\right)^{-1}\sqrt{\mathbf{C}}_{k}(\underline{\mathbf{X}}),\ \ \ k=1,\ldots,K.
\end{equation}
Substituting $\{\boldsymbol{\Gamma}_{i}\}$ to the objective in problem
\eqref{Matrix FP: sum-of-log}, we have 
\begin{equation}
\begin{aligned} & \quad h\left(\mathbf{X},\{\boldsymbol{\Gamma}_{i}\}\right)\\
 & =\sum_{k=1}^{K}\omega_{k}\Biggl(\log\det\left(\mathbf{I}+\sqrt{\mathbf{C}}_{k}^{\mathsf{H}}(\underline{\mathbf{X}})\left(\mathbf{D}_{k}(\underline{\mathbf{X}})\right)^{-1}\sqrt{\mathbf{C}}_{k}(\underline{\mathbf{X}})\right)\\
 & \quad+\mathrm{tr}\Bigl(\mathbf{I}-\left(\mathbf{I}+\sqrt{\mathbf{C}}_{k}^{\mathsf{H}}(\underline{\mathbf{X}})\left(\mathbf{D}_{k}(\underline{\mathbf{X}})\right)^{-1}\sqrt{\mathbf{C}}_{k}(\underline{\mathbf{X}})\right)\cdot\\
 & \quad\left(\mathbf{I}-\sqrt{\mathbf{C}}_{k}^{\mathsf{H}}(\mathbf{x})\left(\mathbf{C}_{k}(\mathbf{x})+\mathbf{D}_{k}(\mathbf{x})\right)^{-1}\sqrt{\mathbf{C}}_{k}(\mathbf{x})\right)\Bigr)\Biggr).
\end{aligned}
\end{equation}
Applying Lemma \ref{1 Transform: MM-log-MIMO} to the objective of
\eqref{Matrix FP: sum-of-log} with $\mathbf{Z}=\mathbf{I}+\sqrt{\mathbf{C}}_{k}^{\mathsf{H}}(\mathbf{x})\left(\mathbf{D}_{k}(\mathbf{x})\right)^{-1}\sqrt{\mathbf{C}}_{k}(\mathbf{x})$,
we also obtain $h\left(\mathbf{X},\{\boldsymbol{\Gamma}_{i}\}\right)$.\footnote{The following result would be necessary for the derivation: the term
$\left(\mathbf{I}-\sqrt{\mathbf{C}}_{k}^{\mathsf{H}}(\mathbf{x})\left(\mathbf{C}_{k}(\mathbf{x})+\mathbf{D}_{k}(\mathbf{x})\right)^{-1}\sqrt{\mathbf{C}}_{k}(\mathbf{x})\right)$
in $h\left(\mathbf{X},\{\boldsymbol{\Gamma}_{i}\}\right)$ simplifies
to $\left(\mathbf{I}+\sqrt{\mathbf{C}}_{k}^{\mathsf{H}}(\mathbf{x})\left(\mathbf{D}_{k}(\mathbf{x})\right)^{-1}\sqrt{\mathbf{C}}_{k}(\mathbf{x})\right)^{-1}$
by using the Woodbury identity.}
\end{IEEEproof}
\begin{prop}
\label{thm:Matrix Quadratic MM} Updating variables $\{\boldsymbol{\Phi}_{i}\}$
in Problem \eqref{Matrix FP: quadratic transformed} by BCA can be
seen as constructing a surrogate function based on Lemma \ref{1 Transform: MM-quadratic-MIMO}
for the objective in \eqref{Matrix FP: sum-of-functions-of-matrix-ratio}
by MM. 
\end{prop}
\begin{IEEEproof}
With $\mathbf{X}$ fixed, $\{\boldsymbol{\Phi}_{i}\}$ is optimally
determined by
\begin{equation}
\boldsymbol{\Phi}_{k}=\left(\mathbf{D}_{k}(\underline{\mathbf{X}})\right)^{-1}\sqrt{\mathbf{C}}_{k}(\underline{\mathbf{X}}),\ \ \ k=1,\ldots,K.
\end{equation}
Substituting $\{\boldsymbol{\Phi}_{i}\}$ to the objective in problem
\eqref{Matrix FP: quadratic transformed}, we have
\begin{equation}
\begin{aligned}h & \left(\mathbf{X},\{\boldsymbol{\Phi}_{i}\}\right)=\sum_{k=1}^{K}f_{k}\Bigl(2\mathrm{Re}\left(\sqrt{\mathbf{C}}_{k}^{\mathsf{H}}(\mathbf{x})\left(\mathbf{D}_{k}(\underline{\mathbf{X}})\right)^{-1}\sqrt{\mathbf{C}}_{k}(\underline{\mathbf{X}})\right)\\
 & -\sqrt{\mathbf{C}}_{k}^{\mathsf{H}}(\underline{\mathbf{X}})\left(\mathbf{D}_{k}(\underline{\mathbf{X}})\right)^{-1}\mathbf{D}_{k}(\mathbf{x})\left(\mathbf{D}_{k}(\underline{\mathbf{X}})\right)^{-1}\sqrt{\mathbf{C}}_{k}(\underline{\mathbf{X}})\Bigr),
\end{aligned}
\end{equation}
which can also be obtained by applying Lemma \ref{1 Transform: MM-quadratic-MIMO}
to the objective in \eqref{Matrix FP: sum-of-functions-of-matrix-ratio}
with $\mathbf{Z}_{1}=\sqrt{\mathbf{C}}_{k}(\mathbf{x})$ and $\mathbf{Z}_{2}=\mathbf{D}_{k}(\mathbf{x})$. 
\end{IEEEproof}

\section{A Novel Algorithm for WSR Maximization\label{Section: analytical solution}}

The beamformer update procedures in WMMSE/WSR-FP and WSR-MM in Sections
\ref{Section: MISO} and \ref{Section: MIMO} necessitate the iterative
adjustment of Lagrangian multipliers, an operation inherently reliant
on the frequent calculation of matrix pseudo-inverses. This nested
iteration scheme introduces a dual-layer loop structure for the algorithms,
imposing a computational complexity that scales cubically with the
number of transmit antennas---a considerable impediment for systems
with extensive antenna arrays. Furthermore, the convergence and monotonicity
traits of these algorithms are intimately tied to the accuracy of
the Lagrangian tuning process, presenting additional challenges. To
overcome the aforementioned issues, we introduce herein a streamlined,
single-loop algorithm with reduced per-iteration computational demands
and assured convergence---the WSR-MM+. Distinguished by its analytical
update expressions for beamformers that eschew matrix inversions,
WSR-MM+ presents a substantial computational advantage. 

\subsection{The Proposed WSR-MM+ Algorithm}

We first discuss the algorithm for MIMO beamforming \eqref{MIMO: WSR Maximization Problem}.
To start with, we introduce a useful result. 
\begin{prop}[\cite{sun2016majorization}]
\label{Lemma: MM quadratic Matrix} Let $(\mathbf{L},\mathbf{M})\in\mathbb{H}^{n}$
such that $\mathbf{M}\succeq\mathbf{L}$. Given $\mathbf{X}^{\mathsf{H}}\mathbf{L}\mathbf{X}$
with $\mathbf{X}\in\mathbb{C}^{n\times m}$, at $\underline{\mathbf{X}}$,
we have 
\begin{equation}
\begin{aligned}\mathrm{tr}\Bigl(\mathbf{X}^{\mathsf{H}}\mathbf{M}\mathbf{X}\Bigr) & \geq\mathrm{tr}\left(\mathbf{X}^{\mathsf{H}}\mathbf{L}\mathbf{X}\right)+2\mathrm{Re}(\mathbf{X}^{\mathsf{H}}(\mathbf{M}-\mathbf{L})\underline{\mathbf{X}})\\
 & \quad+\mathrm{tr}\Bigl(\underline{\mathbf{X}}^{\mathsf{H}}(\mathbf{L}-\mathbf{M})\underline{\mathbf{X}}\Bigr).
\end{aligned}
\label{eq:MM quadratic matrix}
\end{equation}
\end{prop}
Based on Proposition \ref{Lemma: MM quadratic Matrix}, we construct
a novel surrogate function for problem \eqref{MIMO: WSR Maximization Problem}.
This construction hinges on the application of the above result to
the quadratic term in $\mathbf{W}_{k}$, for $k=1,\ldots,K$, encapsulated
within $\ell$ \eqref{eq: f_bar MIMO}, specifically, $-\mathrm{tr}\bigl(\mathbf{W}_{k}^{\mathsf{H}}\sum_{j=1}^{K}\omega_{j}\mathbf{H}_{j,k}^{\mathsf{H}}\mathbf{A}_{j}\mathbf{H}_{j,k}\mathbf{W}_{k}\bigr)$.
Setting $\eta_{k}\geq\lambda_{\max}\left(\sum_{j=1}^{K}\omega_{j}\mathbf{H}_{j,k}^{\mathsf{H}}\mathbf{A}_{j}\mathbf{H}_{j,k}\right)$
we arrive at the surrogate function $\ell^{\prime}$ in \eqref{eq:MM+ matrix}.\footnote{Considering that $\lambda_{\max}\left(\sum_{j=1}^{K}\omega_{j}\mathbf{H}_{j,k}^{\mathsf{H}}\mathbf{A}_{j}\mathbf{H}_{j,k}\right)$
is bounded by the Frobenius norm of $\sum_{j=1}^{K}\omega_{j}\mathbf{H}_{j,k}^{\mathsf{H}}\mathbf{A}_{j}\mathbf{H}_{j,k}$,
and by extension, a computationally efficient choice for $\eta_{k}$
is given by $\sum_{j=1}^{K}\omega_{j}\left\Vert \mathbf{A}_{j}\right\Vert \left\Vert \mathbf{H}_{j,k}\right\Vert ^{2}$.}
\begin{figure*}[!t]
\vspace{-20bp}
\begin{equation}
\begin{aligned}\ell^{\prime}(\{\mathbf{W}_{i}\},\{\underline{\mathbf{W}_{i}}\}) & =-\sum_{k=1}^{K}\mathrm{tr}\Bigl(\eta_{k}\mathbf{W}_{k}^{\mathsf{H}}\mathbf{W}_{k}+2\mathrm{Re}\Bigl(\mathbf{W}_{k}^{\mathsf{H}}(\sum_{j=1}^{K}\omega_{j}\mathbf{H}_{j,k}^{\mathsf{H}}\mathbf{A}_{j}\mathbf{H}_{j,k}-\eta_{k}\mathbf{I})\ensuremath{\underline{\mathbf{W}}_{k}}\Bigr)+\underline{\mathbf{W}}_{k}^{\mathsf{H}}(\eta_{k}\mathbf{I}-\sum_{j=1}^{K}\omega_{j}\mathbf{H}_{j,k}^{\mathsf{H}}\mathbf{A}_{j}\mathbf{H}_{j,k})\underline{\mathbf{W}}_{k}\Bigr)\\
 & \quad+\sum_{k=1}^{K}\omega_{k}2{\rm {\rm Re}}\bigl({\rm tr}(\mathbf{B}_{k}\mathbf{H}_{k,k}\mathbf{W}_{k})\bigr)+\sum_{k=1}^{K}\omega_{k}\left(\underline{\mathsf{R}}_{k}-{\rm tr}(\underline{\mathbf{W}}_{k}^{\mathsf{H}}\mathbf{H}_{k,k}^{\mathsf{H}}\underline{\mathbf{F}}_{k}^{-1}\mathbf{H}_{k,k}\underline{\mathbf{W}}_{k})-\mathrm{tr}(\mathbf{A}_{k})\sigma_{k}^{2}\right).
\end{aligned}
\label{eq:MM+ matrix}
\end{equation}
\rule[0.5ex]{1\textwidth}{1pt}
\end{figure*}
 Given $\ell^{\prime}$ is a surrogate function of $\ell$ and hence
a surrogate of $f$ in problem \eqref{MIMO: WSR Maximization Problem},
the MM principle dictates that we can iteratively tackle the following
optimization:
\begin{equation}
\begin{aligned} & \underset{\{\mathbf{W}_{i}\}\in\mathcal{W}}{\text{minimize}} &  & \sum_{k=1}^{K}\eta_{k}\left\Vert \mathbf{W}_{k}-\mathbf{Q}_{k}\right\Vert ^{2},\end{aligned}
\label{eq:MIMO-MM+}
\end{equation}
with
\begin{equation}
\mathbf{Q}_{k}=\eta_{k}^{-1}\Bigl(\omega_{k}\mathbf{H}_{k,k}^{\mathsf{H}}\mathbf{B}_{k}^{\mathsf{H}}-\bigl(\sum_{j=1}^{K}\omega_{j}\mathbf{H}_{j,k}^{\mathsf{H}}\mathbf{A}_{j}\mathbf{H}_{j,k}-\eta_{k}\mathbf{I}\bigr)\ensuremath{\underline{\mathbf{W}}_{k}}\Bigr).\label{eq:Q tilde with Q in MM}
\end{equation}
Problem \eqref{eq:MIMO-MM+} has an analytical solution as follows:

\noindent %
\noindent\fbox{\begin{minipage}[t]{1\columnwidth - 2\fboxsep - 2\fboxrule}%
\begin{center}
\labelText{\textbf{A7}}{label:MIMO-MM+}: A WSR-MM+ algorithm for
MIMO beamforming
\[
\mathbf{W}_{k}=\mathbf{Q}_{k}\min\left\{ \sqrt{\frac{P}{\left\Vert \mathbf{Q}_{k}\right\Vert _{\mathsf{}}^{2}}},1\right\} .
\]
\par\end{center}%
\end{minipage}}
\begin{rem}
It is noteworthy that WSR-MM+ is compatible with the prevalent per-antenna
power constraints \cite{yu2007transmitter}, in which case analytical
solutions are also attainable in the subproblems. However, a detailed
exposition is precluded due to space limitations.
\end{rem}
Compared to WSR-MM, WSR-MM+ obviates the need for Lagrange multiplier
tuning and repetitive matrix (pseudo-)inversions. Consequently, the
efficiency of the overall algorithm could be enhanced. WSR-MM+ for
MISO beamforming \eqref{0 WSR Maximization} can be derived similarly
via Proposition \ref{Lemma: MM quadratic Matrix} by restricting $\mathbf{X}$
to be a column vector. Applying Proposition \ref{Lemma: MM quadratic Matrix}
to the quadratic term in $\ell$ \eqref{eq:f_bar MISO}, with $\eta\geq\lambda_{\max}\left(\sum_{j=1}^{K}\omega_{j}a_{j}\mathbf{h}_{j}\mathbf{h}_{j}^{\mathsf{H}}\right)$
(A practical selection for $\eta$ mirrors that in the MIMO scenario,
being $\sum_{j=1}^{K}\omega_{j}a_{j}\left\Vert \mathbf{h}_{j}\right\Vert ^{2}$.)
gives rise to the surrogate function $\ell^{\prime}$ in \eqref{eq:MM+ objective in MISO-1}.
\begin{figure*}[!t]
\vspace{-10bp}
\begin{equation}
\begin{aligned}\ell^{\prime}(\{\mathbf{w}_{i}\},\{\underline{\mathbf{w}_{i}}\}) & =-\sum_{k=1}^{K}\Bigl(\phi\mathbf{w}_{k}^{\mathsf{H}}\mathbf{w}_{k}+2\mathrm{Re}\bigl(\underline{\mathbf{w}}_{k}^{\mathsf{H}}(\sum_{j=1}^{K}\omega_{j}a_{j}\mathbf{h}_{j}\mathbf{h}_{j}^{\mathsf{H}}-\phi\mathbf{I})\mathbf{w}_{k}\bigr)+\underline{\mathbf{w}}_{k}^{\mathsf{H}}(\phi\mathbf{I}-\sum_{j=1}^{K}\omega_{j}a_{j}\mathbf{h}_{j}\mathbf{h}_{j}^{\mathsf{H}})\underline{\mathbf{w}}_{k}\Bigr)\\
 & \quad+\sum_{k=1}^{K}\omega_{k}2\mathrm{Re}(b_{k}\mathbf{h}_{k}^{\mathsf{H}}\mathbf{w}_{k})+\sum_{k=1}^{K}\omega_{k}(\underline{\mathsf{R}}_{k}-\underline{\mathsf{SINR}}_{k}-a_{k}\sigma_{k}^{2})
\end{aligned}
\label{eq:MM+ objective in MISO-1}
\end{equation}
\rule[0.5ex]{1\textwidth}{1pt}

\end{figure*}
 Define
\begin{equation}
\mathbf{q}_{k}=\eta^{-1}\Bigl(\omega_{k}b_{k}^{*}\mathbf{h}_{k}-(\sum_{j=1}^{K}\omega_{j}a_{j}\mathbf{h}_{j}\mathbf{h}_{j}^{\mathsf{H}}-\phi\mathbf{I})\underline{\mathbf{w}}_{k}\Bigr),\label{eq:q tilde with q in MM}
\end{equation}
WSR-MM+ suffices to iteratively solve the following problem
\begin{equation}
\begin{aligned} & \underset{\{\mathbf{w}_{i}\}\in\mathcal{W}}{\text{minimize}} &  & \sum_{k=1}^{K}\eta\left\Vert \mathbf{w}_{k}-\mathbf{q}_{k}\right\Vert ^{2},\end{aligned}
\label{eq:MISO-MM+}
\end{equation}
which has a closed-form solution. WSR-MM+ in the MISO scenario is
as follows:

\noindent\fbox{\begin{minipage}[t]{1\columnwidth - 2\fboxsep - 2\fboxrule}%
\begin{center}
\labelText{\textbf{A8}}{label:MISO-MM+}: A WSR-MM+ algorithm for
MISO beamforming
\[
\mathbf{w}_{k}=\mathbf{q}_{k}\min\left\{ \sqrt{\frac{P}{\sum_{k=1}^{K}\left\Vert \mathbf{q}_{k}\right\Vert ^{2}}},1\right\} .
\]
\par\end{center}%
\end{minipage}}
\begin{rem}
The efficiency of the WSR-MM+ algorithm may be further improved through
methods including the extrapolation technique, as described in Section
VII-E of \cite{zhang2021weighted}. Although an extensive exploration
of these techniques is beyond the scope of this paper, readers interested
in a comprehensive discussion are encouraged to refer to \cite{phan2023inertial}.
\end{rem}

\subsection{Interpreting WSR-MM+ as BCA}

In Section \ref{Section: connection}, it is demonstrated that the
updating of auxiliary variables through equivalent transforms within
WSR-FP can be seen as procedures for the creation of surrogate functions
in WSR-MM. Conversely, these equivalent transforms also serve to parameterize
certain intermediary variables within WSR-MM. This perspective allows
us to interpret the newly proposed WSR-MM+ as a BCA algorithm. To
substantiate this perspective, we introduce a novel transform that
aligns with the principles outlined in Proposition \ref{Lemma: MM quadratic Matrix}. 

\begin{figure*}[!t]
\vspace{-10bp}

\begin{equation}
\begin{aligned}\hspace{-1cm} & \underset{\{\boldsymbol{\Gamma}_{i}\},\{\boldsymbol{\Phi}_{i}\},\left(\{\mathbf{W}_{i}\},\{\mathbf{T}_{i}\}\right)\in\mathcal{W}}{\text{maximize}} &  & \hspace{-0.5cm}-\sum_{k=1}^{K}\mathrm{tr}\biggl(\eta_{k}\mathbf{W}_{k}^{\mathsf{H}}\mathbf{W}_{k}+2\mathrm{Re}\Bigl(\mathbf{W}_{k}^{\mathsf{H}}\Bigl(\sum_{j=1}^{K}\omega_{j}\mathbf{H}_{j,k}^{\mathsf{H}}\boldsymbol{\Phi}_{j}(\mathbf{I}+\boldsymbol{\Gamma}_{j})\boldsymbol{\Phi}_{j}^{\mathsf{H}}\mathbf{H}_{j,k}-\eta_{k}\mathbf{I}\Bigr)\mathbf{T}_{k}\Bigr)\\
 &  &  & \hspace{-0.5cm}+\mathbf{T}_{k}^{\mathsf{H}}\Bigl(\eta_{k}\mathbf{I}-\sum_{j=1}^{K}\omega_{j}\mathbf{H}_{j,k}^{\mathsf{H}}\boldsymbol{\Phi}_{j}(\mathbf{I}+\boldsymbol{\Gamma}_{j})\boldsymbol{\Phi}_{j}^{\mathsf{H}}\mathbf{H}_{j,k}\Bigr)\mathbf{T}_{k}\biggr)+\sum_{k=1}^{K}\Biggl(2\mathrm{tr}\Bigl((\mathbf{I}+\boldsymbol{\Gamma}_{k})\mathrm{Re}(\sqrt{\omega_{k}}\mathbf{H}_{k,k}\mathbf{W}_{k}\boldsymbol{\Phi}_{k})\Bigr)\\
 &  &  & \hspace{-0.5cm}+\sum_{k=1}^{K}\omega_{k}\bigl(\log\det(\mathbf{I}+\boldsymbol{\Gamma}_{k})-\mathrm{tr}(\boldsymbol{\Gamma}_{k})\bigr)\Biggr)-\sum_{k=1}^{K}\sigma_{k}^{2}\mathrm{tr}\bigl((\mathbf{I}+\boldsymbol{\Gamma}_{k})\boldsymbol{\Phi}_{k}^{\mathsf{H}}\boldsymbol{\Phi}_{k}\bigr).
\end{aligned}
\label{eq:FP+ Problem}
\end{equation}
\rule[0.5ex]{1\textwidth}{1pt}
\begin{equation}
\begin{aligned} & \underset{\{\gamma_{i}\},\{\phi_{i}\},\left(\{\mathbf{w}_{i}\},\{\mathbf{t}_{i}\}\right)\in\mathcal{W}}{\text{maximize}} &  & -\sum_{k=1}^{K}\Bigl(\eta\mathbf{w}_{k}^{\mathsf{H}}\mathbf{w}_{k}+2\mathrm{Re}\bigl(\mathbf{w}_{k}^{\mathsf{H}}(\sum_{j=1}^{K}|\phi_{j}|^{2}\mathbf{h}_{j}\mathbf{h}_{j}^{\mathsf{H}}-\eta\mathbf{I})\mathbf{t}_{k}\bigr)+\mathbf{t}_{k}^{\mathsf{H}}(\eta\mathbf{I}-\sum_{j=1}^{K}|\phi_{j}|^{2}\mathbf{h}_{j}\mathbf{h}_{j}^{\mathsf{H}})\mathbf{t}_{k}\Bigr)\\
 &  &  & +\sum_{k=1}^{K}2\mathrm{Re}\bigl(\phi_{k}^{*}\sqrt{\omega_{k}(1+\gamma_{k})}\mathbf{h}_{k}^{\mathsf{H}}\mathbf{w}_{k}\bigr)+\sum_{k=1}^{K}\omega_{k}\left(\log\left(1+\gamma_{k}\right)-\gamma_{k}\right)-\sum_{k=1}^{K}|\phi_{k}|^{2}\sigma_{k}^{2}.
\end{aligned}
\label{eq:FP+ objective in MISO}
\end{equation}
\rule[0.5ex]{1\textwidth}{1pt}
\end{figure*}

\begin{cor}[Matrix non-homogeneous transform]
\label{ET: induced from MM quadratic matrix} Let $\mathbf{L},\mathbf{M}\in\mathbb{H}^{n}$
such that $\mathbf{M}\succeq\mathbf{L}$. Problem 
\begin{equation}
\underset{\mathbf{X}\in\mathcal{X}}{\text{minimize}}\ \ \mathrm{tr}\left(\mathbf{X}^{\mathsf{H}}\mathbf{L}\mathbf{X}\right),
\end{equation}
where ${\cal X}$ denotes the constraint for $\mathbf{X}$, is equivalent
to 
\begin{equation}
\underset{(\mathbf{X},\mathbf{Z})\in\mathcal{X}}{\text{minimize}}\ \ \mathrm{tr}\Bigl(\mathbf{X}^{\mathsf{H}}\mathbf{M}\mathbf{X}+2\mathrm{Re}(\mathbf{X}^{\mathsf{H}}(\mathbf{L}-\mathbf{M})\mathbf{Z})+\mathbf{Z}^{\mathsf{H}}(\mathbf{M}-\mathbf{L})\mathbf{Z}\Bigr),\label{eq: matrix NHT}
\end{equation}
in the sense that they attain the identical optimal solution with
the identical optimal objective value. 
\end{cor}
\begin{IEEEproof}
$\mathbf{Z}$ is optimally determined by deriving the first order
optimality condition of \eqref{eq: matrix NHT}. Substituting the
solution into the transformed objective recovers the original one. 
\end{IEEEproof}
Invoking Corollary \ref{ET: induced from MM quadratic matrix}, we
obtain a BCA-based algorithm, termed WSR-FP+, emerging as a counterpart
of WSR-MM+. Applying Corollary \ref{ET: induced from MM quadratic matrix}
to Problem \eqref{MIMO FP: Final Problem} using $\eta_{k}\geq\lambda_{\max}\left(\sum_{j=1}^{K}\omega_{j}\mathbf{H}_{j,k}^{\mathsf{H}}\boldsymbol{\Phi}_{j}\left(\mathbf{I}+\boldsymbol{\Gamma}_{j}\right)\boldsymbol{\Phi}_{j}^{\mathsf{H}}\mathbf{H}_{j,k}\right)$,
we get an equivalent problem presented in \eqref{eq:FP+ Problem}.
Solving \eqref{eq:FP+ Problem} through BCA leads the Algorithm \nameref{label:MIMO-FP+}.

\noindent\fbox{\begin{minipage}[t]{1\columnwidth - 2\fboxsep - 2\fboxrule}%
\begin{center}
\labelText{\textbf{A9}}{label:MIMO-FP+}: A WSR-FP+ algorithm for
MIMO beamforming 
\begin{align*}
\text{S1: } & \mathbf{T}_{k}=\underline{\mathbf{W}}_{k},\\
\text{S2: } & \boldsymbol{\Phi}_{k}=\left(\underline{\mathbf{F}}_{k}+\mathbf{H}_{k,k}\underline{\mathbf{W}}_{k}\underline{\mathbf{W}}_{k}^{\mathsf{H}}\mathbf{H}_{k,k}^{\mathsf{H}}\right)^{-1}\sqrt{\omega_{k}}\mathbf{H}_{k,k}\underline{\mathbf{W}}_{k},\\
\text{S3: } & \boldsymbol{\Gamma}_{k}=\underline{\mathbf{W}}_{k}^{\mathsf{H}}\mathbf{H}_{k,k}^{\mathsf{H}}\underline{\mathbf{F}}_{k}^{-1}\mathbf{H}_{k,k}\underline{\mathbf{W}}_{k},\\
\text{S4: } & \mathbf{W}_{k}=\mathbf{Q}_{k}\min\left\{ \sqrt{\frac{P}{\left\Vert \mathbf{Q}_{k}\right\Vert _{\mathsf{}}^{2}}},1\right\} .
\end{align*}
\par\end{center}%
\end{minipage}}

It can be verified that $\mathbf{Q}_{k}$ in \nameref{label:MIMO-FP+}-S4
is given by \eqref{eq:Q tilde with Q in MM}. In a similar vein,
WSR-FP+ can be used for MISO beamforming \eqref{0 WSR Maximization}.
Based on Corollary \ref{ET: induced from MM quadratic matrix}, choosing
$\eta\geq\lambda_{\max}\left(\sum_{j=1}^{K}|\underline{y}_{j}|^{2}\mathbf{h}_{j}\mathbf{h}_{j}^{\mathsf{H}}\right)$
leads to Problem \eqref{eq:FP+ objective in MISO}. Using BCA to solve
this problem, we obtain Algorithm \nameref{label:MISO-FP+}, where
$\mathbf{q}_{k}$ appeared in \nameref{label:MISO-FP+}-S4 is given
in \eqref{eq:q tilde with q in MM}.

\noindent\fbox{\begin{minipage}[t]{1\columnwidth - 2\fboxsep - 2\fboxrule}%
\begin{center}
\labelText{\textbf{A10}}{label:MISO-FP+}: A WSR-FP+ algorithm for
MISO beamforming 
\begin{align*}
\text{S1: } & \mathbf{t}_{k}=\underline{\mathbf{w}}_{k},\\
\text{S2: } & \gamma_{k}=\underline{\mathsf{SINR}}_{k},\\
\text{S3: } & y_{k}=\frac{\sqrt{\omega_{k}\left(1+\underline{\gamma}_{k}\right)}\mathbf{h}_{k}^{\mathsf{H}}\underline{\mathbf{w}}_{k}}{\sum_{j=1}^{K}\bigl|\mathbf{h}_{k}^{\mathsf{H}}\underline{\mathbf{w}}_{j}\bigr|^{2}+\sigma^{2}},\\
\text{S4: } & \mathbf{w}_{k}=\mathbf{q}_{k}\min\left\{ \sqrt{\frac{P}{\sum_{k=1}^{K}\left\Vert \mathbf{q}_{k}\right\Vert ^{2}}},1\right\} .
\end{align*}
\par\end{center}%
\end{minipage}}

\subsection{Interpreting WSR-MM+ as Projected Gradient Ascent \label{subsec:PGA-MIMO}}

\begin{figure*}[!t]
\begin{equation}
\begin{aligned}\frac{\partial f\left(\{\mathbf{W}_{i}\}\right)}{\partial\mathbf{W}_{k}} & =2\omega_{k}\mathbf{H}_{k,k}^{\mathsf{H}}\mathbf{F}_{k}^{-1}\mathbf{H}_{k,k}\mathbf{W}_{k}\left(\mathbf{I}+\mathbf{W}_{k}^{\mathsf{H}}\mathbf{H}_{k,k}^{\mathsf{H}}\mathbf{F}_{k}^{-1}\mathbf{H}_{k,k}\mathbf{W}_{k}\right)^{-1}\\
 & \quad-2\sum_{i=1,i\neq k}^{K}\omega_{i}\mathbf{H}_{i,k}^{\mathsf{H}}\mathbf{F}_{i}^{-1}\mathbf{H}_{i,i}\mathbf{W}_{i}\left(\mathbf{I}+\mathbf{W}_{i}^{\mathsf{H}}\mathbf{H}_{i,i}^{\mathsf{H}}\mathbf{F}_{i}^{-1}\mathbf{H}_{i,i}\mathbf{W}_{i}\right)^{-1}\mathbf{W}_{i}^{\mathsf{H}}\mathbf{H}_{i,i}^{\mathsf{H}}\mathbf{F}_{i}^{-1}\mathbf{H}_{i,k}\mathbf{W}_{k}\\
 & =2\omega_{k}\mathbf{H}_{k,k}^{\mathsf{H}}\mathbf{F}_{k}^{-1}\mathbf{H}_{k,k}\mathbf{W}_{k}-2\sum_{i=1}^{K}\omega_{i}\mathbf{H}_{i,k}^{\mathsf{H}}\mathbf{F}_{i}^{-1}\mathbf{H}_{i,i}\mathbf{W}_{i}\left(\mathbf{I}+\mathbf{W}_{i}^{\mathsf{H}}\mathbf{H}_{i,i}^{\mathsf{H}}\mathbf{F}_{i}^{-1}\mathbf{H}_{i,i}\mathbf{W}_{i}\right)^{-1}\mathbf{W}_{i}^{\mathsf{H}}\mathbf{H}_{i,i}^{\mathsf{H}}\mathbf{F}_{i}^{-1}\mathbf{H}_{i,k}\mathbf{W}_{k}\\
 & =2\omega_{k}\mathbf{H}_{k,k}^{\mathsf{H}}\mathbf{F}_{k}^{-1}\mathbf{H}_{k,k}\mathbf{W}_{k}-2\sum_{i=1}^{K}\omega_{i}\mathbf{H}_{i,k}^{\mathsf{H}}(\mathbf{F}_{i}+\mathbf{H}_{i,i}\mathbf{W}_{i}\mathbf{W}_{i}^{\mathsf{H}}\mathbf{H}_{i,i}^{\mathsf{H}})^{-1}\mathbf{H}_{i,i}\mathbf{W}_{i}\mathbf{W}_{i}^{\mathsf{H}}\mathbf{H}_{i,i}^{\mathsf{H}}\mathbf{F}_{i}^{-1}\mathbf{H}_{i,k}\mathbf{W}_{k}.
\end{aligned}
\label{eq:derivative}
\end{equation}
\rule[0.5ex]{1\textwidth}{1pt}
\begin{equation}
\frac{\partial f\left(\{\mathbf{w}_{i}\}\right)}{\partial\mathbf{w}_{k}}=\frac{2\omega_{k}\mathbf{h}_{k}\mathbf{h}_{k}^{\mathsf{H}}\mathbf{w}_{k}}{\sum_{j=1,j\neq k}^{K}\bigl|\mathbf{h}_{k}^{\mathsf{H}}\mathbf{w}_{j}\bigr|^{2}+\sigma_{k}^{2}}-\sum_{k=1}^{K}\frac{\omega_{k}\bigl|\mathbf{h}_{k}^{\mathsf{H}}\mathbf{w}_{k}\bigr|^{2}}{\sum_{j=1,j\neq k}^{K}\bigl|\mathbf{h}_{k}^{\mathsf{H}}\mathbf{w}_{j}\bigr|^{2}+\sigma_{k}^{2}}\frac{2\mathbf{h}_{k}\mathbf{h}_{k}^{\mathsf{H}}\mathbf{w}_{k}}{\sum_{j=1}^{K}\bigl|\mathbf{h}_{k}^{\mathsf{H}}\mathbf{w}_{j}\bigr|^{2}+\sigma_{k}^{2}}.\label{eq:derivative scalar}
\end{equation}
\rule[0.5ex]{1\textwidth}{1pt}
\end{figure*}
In deriving WSR-MM+, the surrogate function $\ell^{\prime}$ is an
isotropic quadratic approximation of the original objective, which
affords WSR-MM+ an intriguing conceptual parallel to the projected
gradient ascent method. Given $f\left(\{\mathbf{W}_{i}\}\right)$
in problem formulation \eqref{MIMO: WSR Maximization Problem}, the
gradient with respect to $\mathbf{W}_{k}$ is given in \eqref{eq:derivative},\footnote{Wirtinger calculus is adopted for differentials of complex variables
\cite{hjorungnes2011complex}.} where the last equality is due to the positive definite identity
\cite{welling2010kalman}: $\mathbf{F}_{i}^{-1}\mathbf{H}_{i,i}\mathbf{W}_{i}\left(\mathbf{I}+\mathbf{W}_{i}^{\mathsf{H}}\mathbf{H}_{i,i}^{\mathsf{H}}\mathbf{F}_{i}^{-1}\mathbf{H}_{i,i}\mathbf{W}_{i}\right)^{-1}=\left(\mathbf{F}_{i}+\mathbf{H}_{i,i}\mathbf{W}_{i}\mathbf{W}_{i}^{\mathsf{H}}\mathbf{H}_{i,i}^{\mathsf{H}}\right)^{-1}\mathbf{H}_{i,i}\mathbf{W}_{i}$.
Then, $\mathbf{Q}_{k}$ in \eqref{eq:Q tilde with Q in MM} can be
rewritten as
\begin{equation}
\begin{aligned}\mathbf{Q}_{k} & =\underline{\mathbf{W}}_{k}+\frac{1}{\eta_{k}}\Bigl(\omega_{k}\mathbf{H}_{k,k}^{\mathsf{H}}\mathbf{B}_{k}^{\mathsf{H}}-\sum_{j=1}^{K}\omega_{j}\mathbf{H}_{j,k}^{\mathsf{H}}\mathbf{A}_{j}\mathbf{H}_{j,k}\underline{\mathbf{W}}_{k}\Bigr)\\
 & =\underline{\mathbf{W}}_{k}+\frac{1}{2\eta_{k}}\frac{\partial f\left(\{\underline{\mathbf{W}}_{i}\}\right)}{\partial\mathbf{W}_{k}}.
\end{aligned}
\end{equation}
Thus, each individual iteration within WSR-MM+, specifically solving
problem \eqref{eq:MISO-MM+}, can be interpreted as executing a step
of projected gradient ascent with step size $\frac{1}{2\eta_{k}}$
(note $\eta_{k}$ depends on $\{\underline{\mathbf{W}}_{i}\}$), formally
described by
\[
\mathbf{W}_{k}=\mathrm{Proj}_{{\cal W}}(\underline{\mathbf{W}}_{k}+\frac{1}{2\eta_{k}}\frac{\partial f\left(\{\underline{\mathbf{W}}_{i}\}\right)}{\partial\mathbf{W}_{k}}),
\]
where $\mathrm{Proj}_{{\cal W}}(\cdot)$ denotes the projection operator
onto set ${\cal W}$. Similar result applies to the MISO case. Considering
the objective function $f\left(\{\mathbf{w}_{i}\}\right)$ in problem
formulation \eqref{0 WSR Maximization}, the gradient with respect
to $\mathbf{w}_{k}$ is provided in \eqref{eq:derivative scalar}.
It follows that $\mathbf{q}_{k}$ in \eqref{eq:q tilde with q in MM}
can be expressed as
\begin{equation}
\begin{aligned}\mathbf{q}_{k} & =\underline{\mathbf{w}}_{k}+\frac{1}{\eta}\Bigl(\omega_{k}b_{k}^{*}\mathbf{h}_{k}-\sum_{j=1}^{K}\omega_{j}a_{j}\mathbf{h}_{j}\mathbf{h}_{j}^{\mathsf{H}}\underline{\mathbf{w}}_{k}\Bigr)\\
 & =\underline{\mathbf{w}}_{k}+\frac{1}{2\eta}\frac{\partial f\left(\{\underline{\mathbf{w}}_{i}\}\right)}{\partial\mathbf{w}_{k}}.
\end{aligned}
\end{equation}
Therefore, WSR-MM+ for MISO beamforming is equivalent to a projected
gradient ascent step with step size $\frac{1}{2\eta}$. 

\section{Per-Iteration Complexity Analysis\label{Section: Complexity}}

In this section, we analyze the complexity of WMMSE (\nameref{label:MISO-WMMSE}
and \nameref{label:MIMO-WMMSE}), WSR-FP (\nameref{label:MISO-FP}
and \nameref{label:MIMO-FP}), WSR-MM (\nameref{label:MISO-MM} and
\nameref{label:MIMO-MM}), WSR-MM+ (\nameref{label:MISO-MM+} and
\nameref{label:MIMO-MM+}), and WSR-FP+ (\nameref{label:MISO-FP+}
and \nameref{label:MIMO-FP+}). For simplicity, we assume $M$ in
the MISO case and $\{M_{i}^{\mathsf{r}}\}$, $\{M_{i}^{\mathsf{t}}\}$,
and $\{M_{i}^{\mathsf{s}}\}$ in the MIMO case are of the same order,
uniformly denoted as $M$. We denote by $I$ the iteration numbers
for searching the Lagrangian multiplier (i.e., \eqref{eq:bisection mu MISO}
or \eqref{eq:bisection mu MIMO}). In the following, we first analyze
the per-iteration computational cost of the WSR-MM and WSR-MM+. 

For MISO case, the complexity of WSR-MM is dominated by the matrix
pseudo-inversions, the vector-vector outer products, and matrix-vector
products. The complexity of WSR-MM+ mainly comes from the multiplication
between the vector-vector outer products and matrix-vector products.
The overall per-iteration complexity of WSR-MM and WSR-MM+ are $\mathcal{O}\left(I\left(KM^{2}+M^{3}\right)\right)$
and $\mathcal{O}\left(KM^{2}\right)$ ($\eta$ is obtained independently
of eigenvalue computations), respectively. For MIMO case, the complexity
of WSR-MM and WSR-MM+ are both dominated by the matrix pseudo-inverse
operation. The overall per-iteration complexity of WSR-MM and WSR-MM+
are $\mathcal{O}\left(I\left(KM^{3}+M^{3}\right)\right)$ and $\mathcal{O}\left(KM^{3}\right)$,
respectively. 

For WMMSE, WSR-FP, and WSR-MM, as discussed in Section \ref{Section: connection},
it becomes apparent that these algorithms exhibit the same order of
complexity. Nonetheless, WSR-MM boasts a distinctive advantage in
that it bypasses the computation of auxiliary variables. Specifically,
WSR-MM can eliminate certain operations that, in the context of WMMSE
and WSR-FP, carry a complexity of $\mathcal{O}\left(K\right)$ for
MISO and $\mathcal{O}\left(KM^{2}\right)$ for MIMO beamforming problems,
respectively. Correspondingly, the analysis extends to WSR-MM+ and
WSR-FP+, wherein both algorithms demonstrate the same order of complexity,
while WSR-MM+ offers additional advantage of obviating the need for
auxiliary variable computations.

\section{Numerical Experiments\label{Section: Experiments}}

In this section, we corroborate our theoretical findings with numerical
experiments performed in MATLAB on a personal computer with a 3.3
GHz Intel Xeon W CPU.\footnote{Code available at $\textsf{https://github.com/zepengzhang/RateMax}$.}

In our evaluation, we investigate both MISO and MIMO systems configured
in a three-dimensional Cartesian setting. For the MISO system, a base
station is placed at $(0,0,10)\mathsf{m}$ serving $K$ users randomly
distributed in a circle centered at $(200,30,0)\mathsf{m}$ with radius
of $10\mathsf{m}$. We assume that the channel fading is frequency
flat and adopts the Rayleigh fading model $\mathbf{h}_{k}=\sqrt{\kappa(d)}\times\mathcal{CN}(\mathbf{0},\mathbf{I})$,
where $\kappa(x)$ represents the distance-dependent path loss. Specifically,
the path loss is computed as $\kappa(d)=T_{0}(\frac{d}{d_{0}})^{-\varrho}$
where the path loss at the reference distance $d_{0}=1\mathsf{m}$
is $T_{0}=-30\mathsf{dB}$ and $\varrho=3.67$ is the path loss exponent.
For the  MIMO system, the channels are modeled analogously to the
MISO case. The transmitters and the receivers are randomly distributed
in two circles centered at $(0,0,10)\mathsf{m}$ and $(200,30,0)\mathsf{m}$,
respectively, both with radius of $10\mathsf{m}$. For both systems,
we have considered the noise power spectrum density of $-169\mathsf{dBm/Hz}$
and the transmission bandwidth of 240$\mathsf{kHz}$. We set $\sigma_{k}^{2}=1$
for $k=1,\ldots,K$, and $P=0\mathsf{dBm}$. We set the algorithm
convergence criterion to be the objective value increment less than
$10^{-6}$. All reported results are averaged over 100 independent
channel realizations.

\begin{figure}
\subfloat[Average WSR versus iteration in the MISO case.\label{fig:MISO-itr}]{\includegraphics[width=0.48\columnwidth]{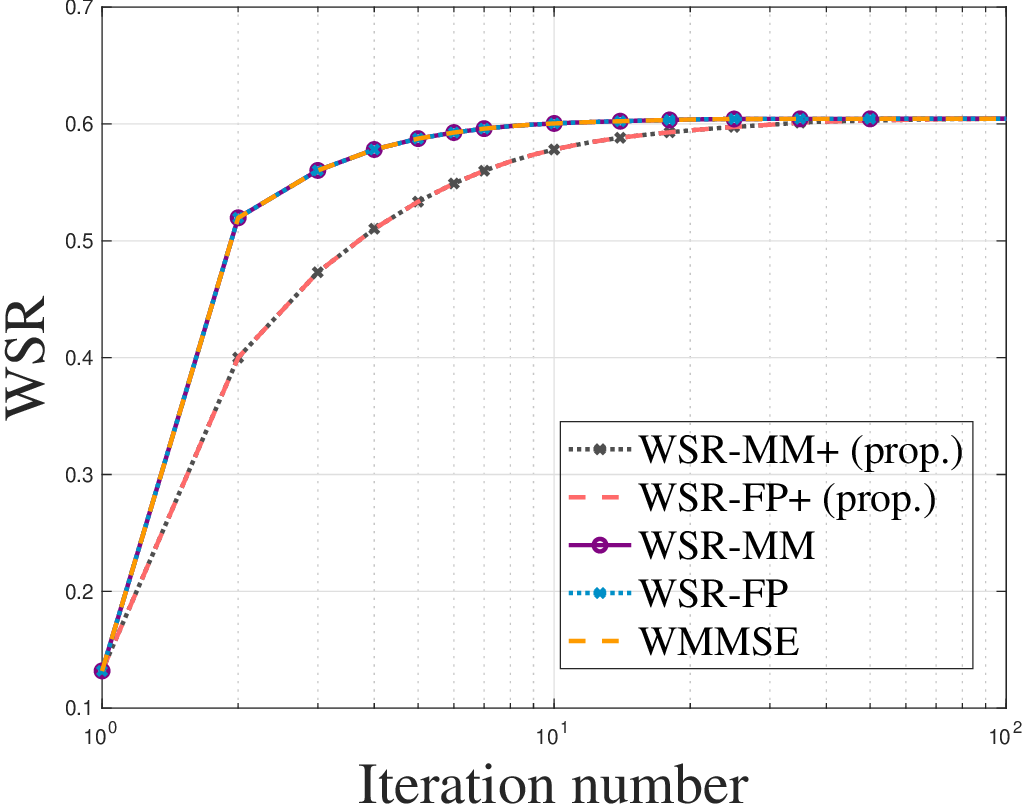}}\subfloat[Average WSR versus CPU time in the MISO case.\label{fig:MISO-time}]{\includegraphics[width=0.48\columnwidth]{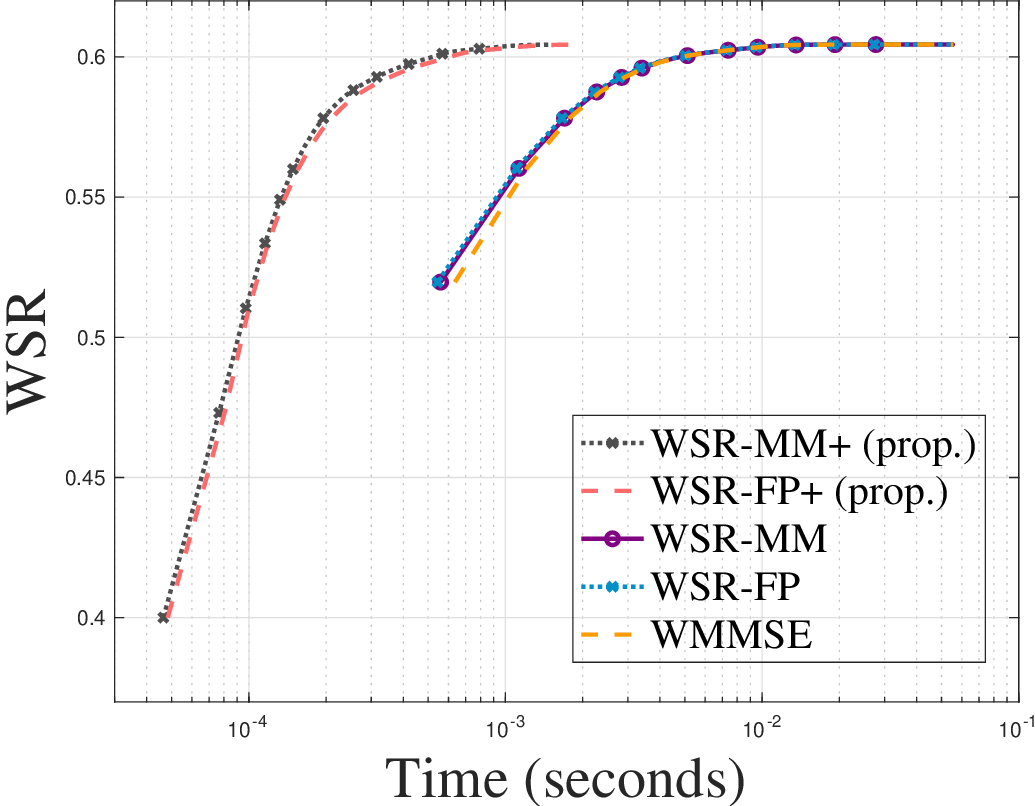}}

\subfloat[Average WSR versus iteration in the MIMO case.\label{fig:MIMO-itr}]{\includegraphics[width=0.48\columnwidth]{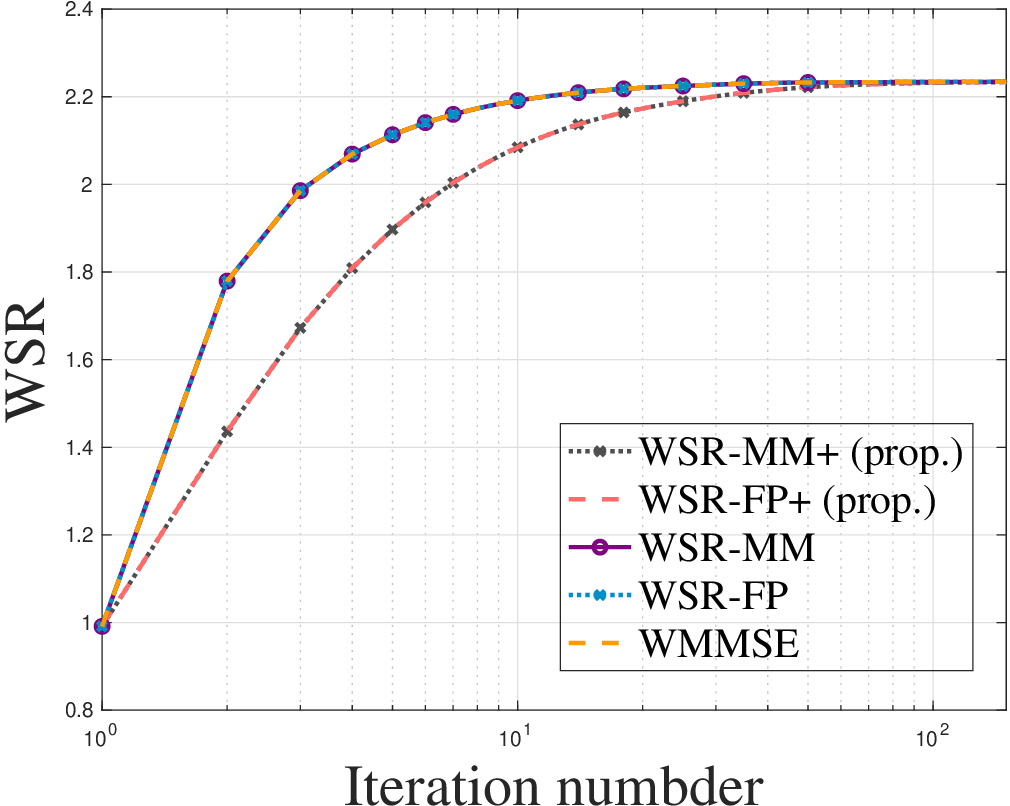}}\subfloat[Average WSR versus CPU time in the MISO case.\label{fig:MIMO-time}]{\includegraphics[width=0.48\columnwidth]{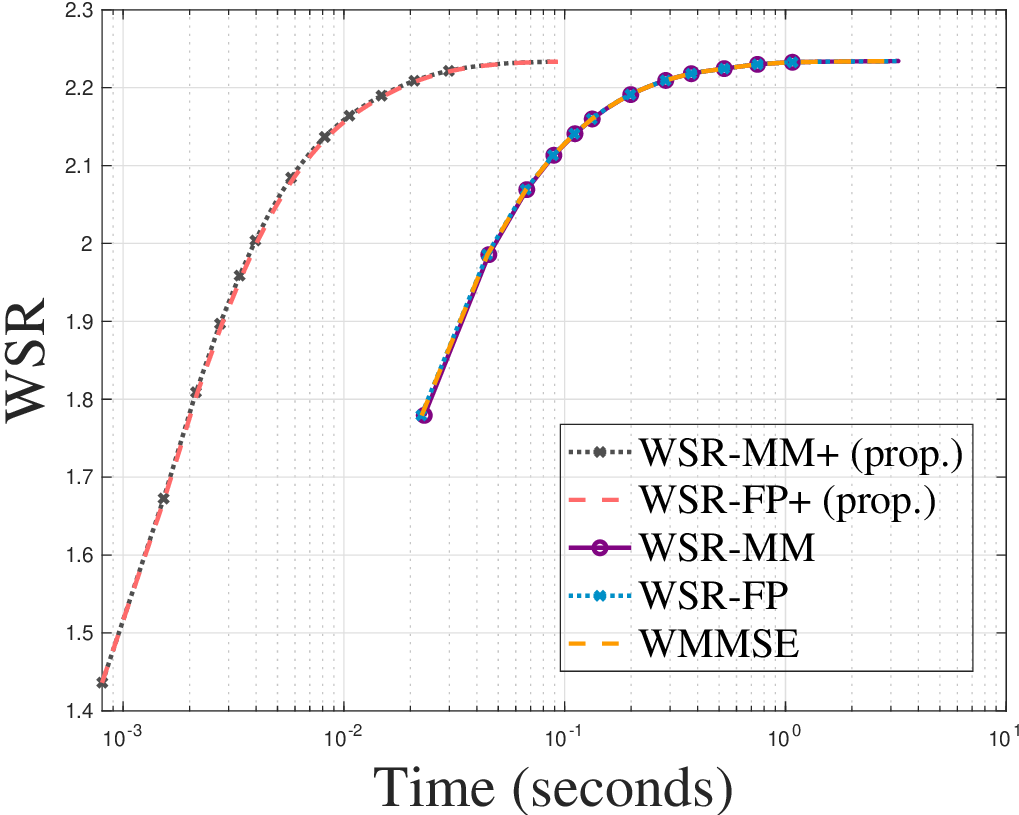}}

\caption{Performance comparison of different algorithms.}
\vspace{-0.4cm}
\end{figure}

We compare the WMMSE, WSR-FP, WSR-MM, WSR-MM+, and WSR-FP+ algorithms
in both a MISO system (with parameters $K=4$ and $M=4$), in which
case we implement \nameref{label:MISO-WMMSE}, \nameref{label:MISO-FP},
\nameref{label:MISO-MM}, \nameref{label:MISO-MM+}, and \nameref{label:MISO-FP+},
and a MIMO system (with parameters $K=4$ and $M_{k}^{\mathsf{r}}=M_{k}^{\mathsf{t}}=M_{k}^{\mathsf{s}}=4$
for $k=1,\ldots,K$), in which case we implement \nameref{label:MIMO-WMMSE},
\nameref{label:MIMO-FP}, \nameref{label:MIMO-MM}, \nameref{label:MIMO-MM+},
and \nameref{label:MIMO-FP+}. Figs. \ref{fig:MISO-itr} and \ref{fig:MISO-time}
illustrate the performance of these algorithms in the MISO setting,
based on the number of iterations and CPU time, respectively. Correspondingly,
Figs. \ref{fig:MIMO-itr} and \ref{fig:MIMO-time} present the analogous
results for the MIMO configuration. From the simulation results, we
can see that the WMMSE, WSR-FP, and WSR-MM algorithms exhibit comparable
performance in both system configurations. This pattern of performance
is consistent with the results obtained for WSR-MM+ and WSR-FP+. Notably,
WSR-MM+ and WSR-FP+ demonstrate a trade-off between the number of
iterations required to achieve convergence and the CPU time consumed.
Despite a higher iteration count, these algorithms benefit from reduced
CPU time, suggesting that the per-iteration analytical solution offers
a significant computational advantage.

\begin{figure}[tbh]
\begin{centering}
\includegraphics[width=0.8\columnwidth]{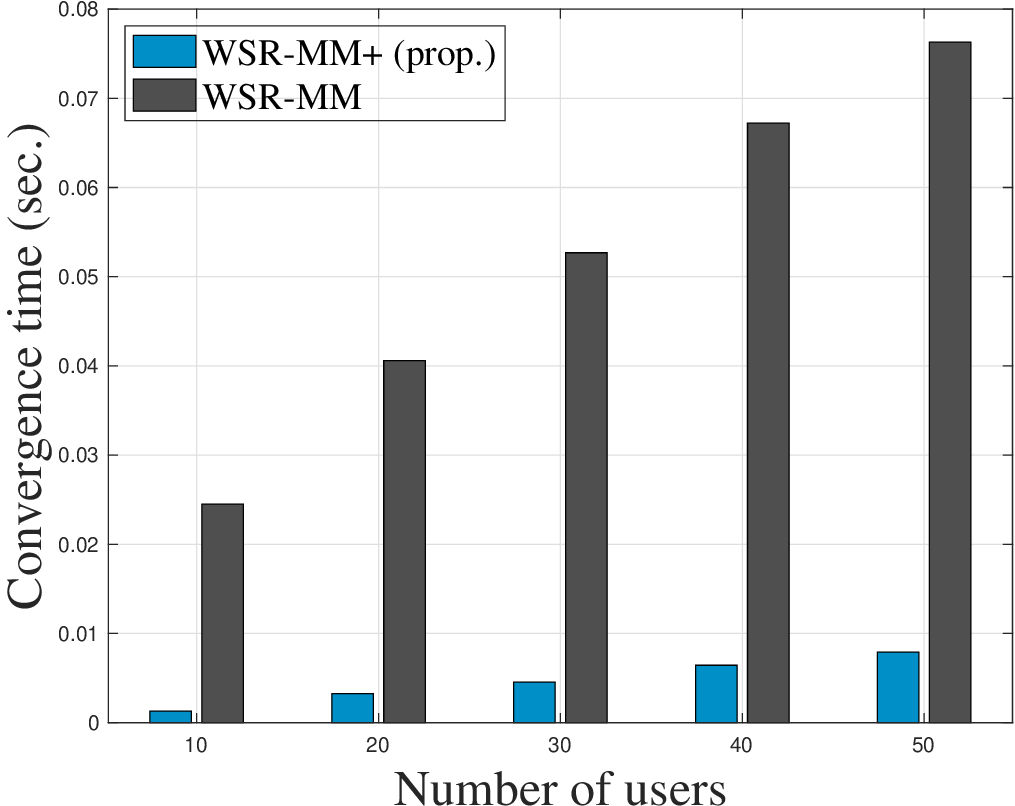}
\par\end{centering}
\caption{Average convergence time versus number of users.\label{fig:Convergence-time-K}}
\vspace{-0.4cm}
\end{figure}

\begin{figure}[tbh]
\centering{}\includegraphics[width=0.8\columnwidth]{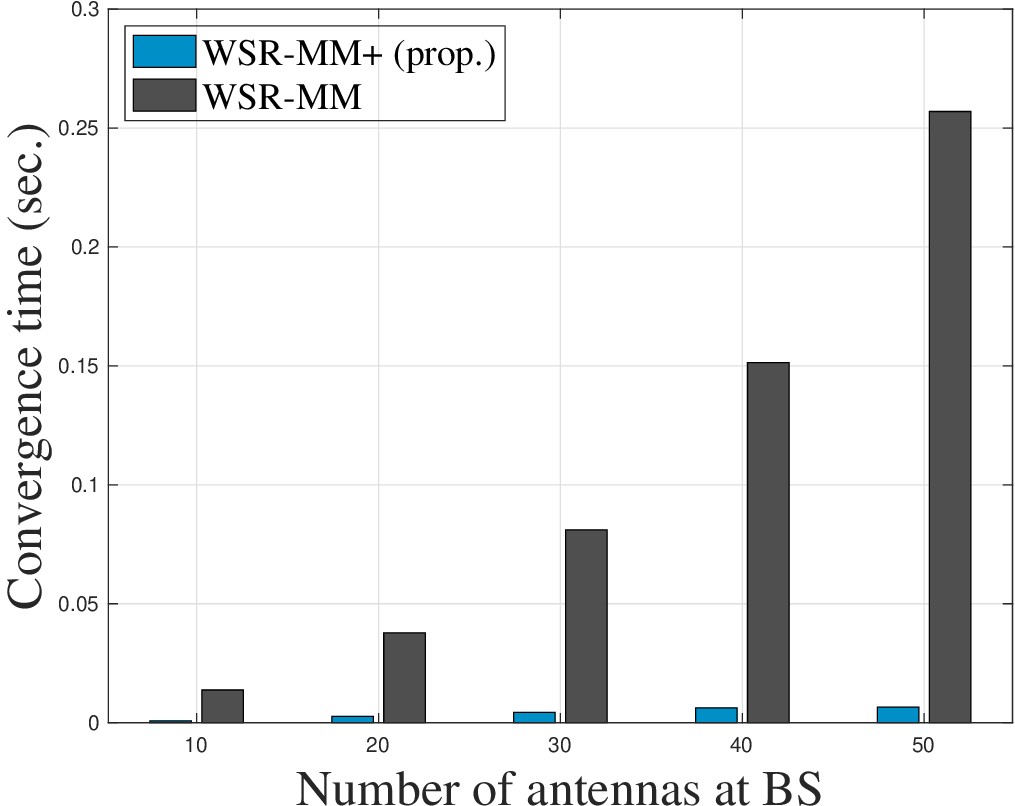}\caption{Average convergence time versus number of antennas at the base station.\label{fig:Convergence-time-M}}
\vspace{-0.2cm}
\end{figure}

To more precisely quantify the performance enhancement offered by
WSR-MM+ over WSR-MM, we investigate their respective convergence times
for WSR maximization in MISO systems varying both number of users
(i.e., $K$) in Fig. \ref{fig:Convergence-time-K} and number of antennas
at the base station (i.e., $M$) in Fig. \ref{fig:Convergence-time-M}.
From the results, we can easily observe that WSR-MM+ significantly
surpasses WSR-MM by a large margin.

\begin{figure}[tbh]
\begin{centering}
\includegraphics[width=0.8\columnwidth]{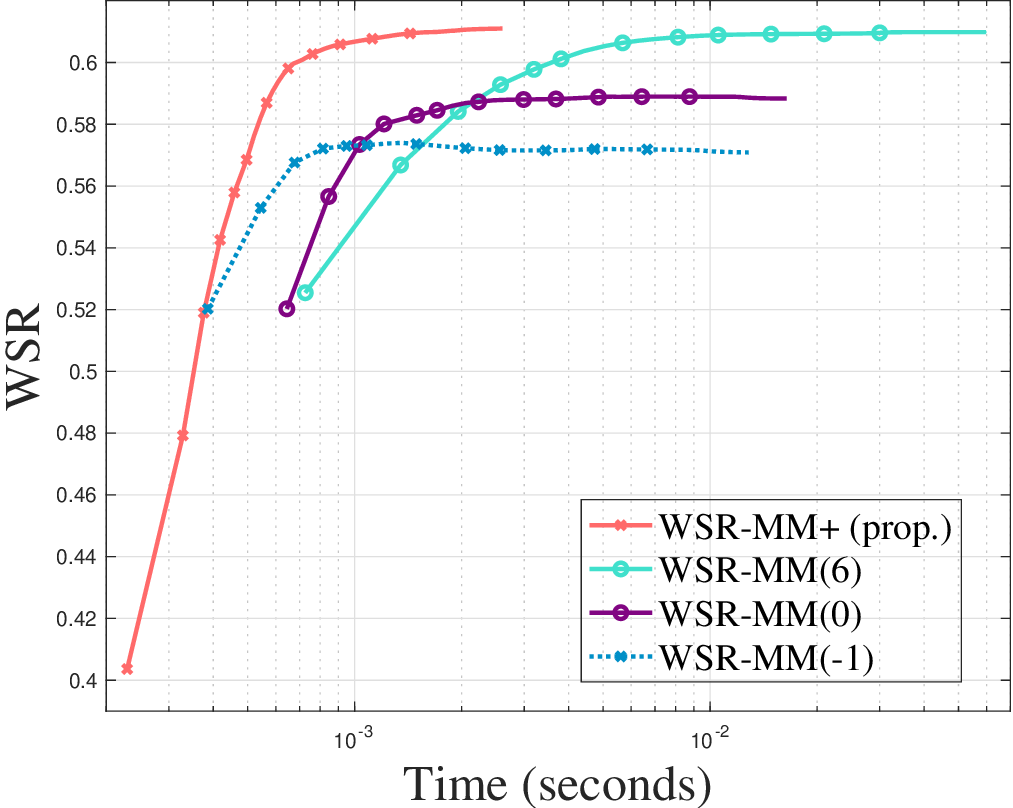}
\par\end{centering}
\caption{Convergence behaviors of WSR-MM and WSR-MM+.\label{fig:threshold}}
\end{figure}

It might be posited that the augmented complexity inherent in WSR-MM
relative to WSR-MM+ is primarily attributed to the one-dimensional
search procedures, as given in \eqref{eq:bisection mu MISO} and \eqref{eq:bisection mu MIMO}.
A conjecture may arise that by loosening the stopping criteria for
these searches, the total computation time for WSR-MM may decrease,
potentially resulting in a more efficient algorithm than WSR-MM+.
To explore this conjecture, Fig. \ref{fig:threshold} provides a numerical
comparison between WSR-MM+ and different versions of of WSR-MM that
are subjected to differing stopping criteria. Specifically, ``WSR-MM$(i)$''
in Fig. \ref{fig:threshold} denotes a WSR-MM algorithm that terminates
the one-dimensional search when the change in the search variable
falls below $2^{-i}$ across successive iterations. Contrary to expectations,
the results indicate that relaxing the stopping criteria fails to
expedite convergence; rather, it can result in a non-monotonic convergence
trajectory or cause the algorithm to settle at suboptimal local minima.

\section{Conclusions and Discussions\label{Section: Conclusion}}

In this work, we have explored the precise connections between the
WMMSE, WSR-FP, and WSR-MM algorithms in addressing the sum-rate maximization
problems. We have shown that specific variants of the WMMSE and WSR-FP
algorithms can be construed within the MM algorithmic paradigm. Furthermore,
we have established that the equivalent transforms employed in WMMSE
and WSR-FP can be regarded as methodologies to construct surrogate
functions within the WSR-MM algorithm. We have also introduced WSR-MM+,
an enhanced algorithm characterized by its analytical update steps.
Complementing this, we have presented WSR-FP+, the BCA equivalent
of WSR-MM+, and revealed its relation to the projected gradient ascent
method. The superiority of WSR-MM+ and WSR-FP+ has been substantiated
through numerical simulations.

Several compelling directions for future research emerge from this
study. A promising one lies in examining the acceleration and convergence
properties of WSR-MM+. Another fertile area for exploration is the
adaptation of WSR-MM+ to more intricate system design challenges,
such as multi-cell communication systems, as well as to rate maximization
problems that encompass multiple design variables beyond just the
transmit beamformers. Additionally, an intriguing prospect involves
the application of algorithm unrolling \cite{monga2021algorithm}
to transform WSR-MM+ into deep neural networks, thereby imbuing certain
algorithmic parameters with learnability. These neural networks after
training could potentially outstrip their iterative predecessors in
terms of convergence and generalization properties.

\bibliographystyle{IEEEtran}
\bibliography{ref}

\begin{thebibliography}{10}
\providecommand{\url}[1]{#1}
\csname url@samestyle\endcsname
\providecommand{\newblock}{\relax}
\providecommand{\bibinfo}[2]{#2}
\providecommand{\BIBentrySTDinterwordspacing}{\spaceskip=0pt\relax}
\providecommand{\BIBentryALTinterwordstretchfactor}{4}
\providecommand{\BIBentryALTinterwordspacing}{\spaceskip=\fontdimen2\font plus
\BIBentryALTinterwordstretchfactor\fontdimen3\font minus
  \fontdimen4\font\relax}
\providecommand{\BIBforeignlanguage}[2]{{%
\expandafter\ifx\csname l@#1\endcsname\relax
\typeout{** WARNING: IEEEtran.bst: No hyphenation pattern has been}%
\typeout{** loaded for the language `#1'. Using the pattern for}%
\typeout{** the default language instead.}%
\else
\language=\csname l@#1\endcsname
\fi
#2}}
\providecommand{\BIBdecl}{\relax}
\BIBdecl

\bibitem{zhang2023enhancing}
Z.~Zhang, Z.~Zhao, and K.~Shen, ``Enhancing the efficiency of wmmse and fp for
  beamforming by minorization-maximization,'' in \emph{IEEE Int. Conf. Acoust.,
  Speech and Signal Process. (ICASSP)}, 2023, pp. 1--5.

\bibitem{weeraddana2012weighted}
P.~C. Weeraddana, M.~Codreanu, M.~Latva-aho, A.~Ephremides, C.~Fischione
  \emph{et~al.}, ``Weighted sum-rate maximization in wireless networks: A
  review,'' \emph{Found. and Trends{\textregistered} in Netw.}, vol.~6, no.
  1--2, pp. 1--163, 2012.

\bibitem{christensen2008weighted}
S.~S. Christensen, R.~Agarwal, E.~De~Carvalho, and J.~M. Cioffi, ``Weighted
  sum-rate maximization using weighted {MMSE} for {MIMO-BC} beamforming
  design,'' \emph{IEEE Trans. Wireless Commun.}, vol.~7, no.~12, pp.
  4792--4799, 2008.

\bibitem{shi2011iteratively}
Q.~Shi, M.~Razaviyayn, Z.-Q. Luo, and C.~He, ``An iteratively weighted mmse
  approach to distributed sum-utility maximization for a {MIMO} interfering
  broadcast channel,'' \emph{IEEE Trans. Signal Process.}, vol.~59, no.~9, pp.
  4331--4340, 2011.

\bibitem{shen2018fractional1}
K.~Shen and W.~Yu, ``Fractional programming for communication systems---{Part
  I: Power} control and beamforming,'' \emph{IEEE Trans. Signal Process.},
  vol.~66, no.~10, pp. 2616--2630, 2018.

\bibitem{zhang2021weighted}
Z.~Zhang and Z.~Zhao, ``Rate maximizations for reconfigurable intelligent
  surface-aided wireless networks: A unified framework via block
  minorization-maximization,'' \emph{arXiv preprint arXiv:2105.02395}, 2021.

\bibitem{luo2008dynamic}
Z.-Q. Luo and S.~Zhang, ``Dynamic spectrum management: Complexity and
  duality,'' \emph{IEEE J. Sel. Topics in Signal Process.}, vol.~2, no.~1, pp.
  57--73, 2008.

\bibitem{liu2010coordinated}
Y.-F. Liu, Y.-H. Dai, and Z.-Q. Luo, ``Coordinated beamforming for {MISO}
  interference channel: Complexity analysis and efficient algorithms,''
  \emph{IEEE Trans. Signal Process.}, vol.~59, no.~3, pp. 1142--1157, 2010.

\bibitem{shen2019optimization}
K.~Shen, W.~Yu, L.~Zhao, and D.~P. Palomar, ``Optimization of {MIMO}
  device-to-device networks via matrix fractional programming: A
  minorization--maximization approach,'' \emph{IEEE/ACM Trans. Netw.}, vol.~27,
  no.~5, pp. 2164--2177, 2019.

\bibitem{zhao2022rethinking}
X.~Zhao, S.~Lu, Q.~Shi, and Z.-Q. Luo, ``Rethinking {WMMSE}: Can its complexity
  scale linearly with the number of {BS} antennas?'' \emph{arXiv preprint
  arXiv:2205.06225}, 2022.

\bibitem{stancu2012fractional}
I.~M. Stancu-Minasian, \emph{Fractional programming: Theory, methods and
  applications}.\hskip 1em plus 0.5em minus 0.4em\relax Springer Sci. \& Bus.
  Media, 2012, vol. 409.

\bibitem{cheung2013achieving}
K.~T.~K. Cheung, S.~Yang, and L.~Hanzo, ``Achieving maximum energy-efficiency
  in multi-relay {OFDMA} cellular networks: A fractional programming
  approach,'' \emph{IEEE Trans. Commun.}, vol.~61, no.~7, pp. 2746--2757, 2013.

\bibitem{zappone2015energy}
A.~Zappone and E.~Jorswieck, ``Energy efficiency in wireless networks via
  fractional programming theory,'' \emph{Found. and Trends in Commun. and Inf.
  Theory}, vol.~11, no. 3-4, pp. 185--396, 2015.

\bibitem{khan2020optimizing}
A.~A. Khan, R.~S. Adve, and W.~Yu, ``Optimizing downlink resource allocation in
  multiuser {MIMO} networks via fractional programming and the {H}ungarian
  algorithm,'' \emph{IEEE Trans. Wireless Commun.}, vol.~19, no.~8, pp.
  5162--5175, 2020.

\bibitem{shen2020enhanced}
K.~Shen, H.~V. Cheng, X.~Chen, Y.~C. Eldar, and W.~Yu, ``Enhanced channel
  estimation in massive {MIMO} via coordinated pilot design,'' \emph{IEEE
  Trans. Commun.}, vol.~68, no.~11, pp. 6872--6885, 2020.

\bibitem{park2021collaborative}
S.-H. Park, S.~Jeong, J.~Na, O.~Simeone, and S.~Shamai, ``Collaborative cloud
  and edge mobile computing in {C-RAN} systems with minimal end-to-end
  latency,'' \emph{IEEE Trans. Signal and Inf. Process. over Netw.}, vol.~7,
  pp. 259--274, 2021.

\bibitem{hunter2004tutorial}
D.~R. Hunter and K.~Lange, ``A tutorial on mm algorithms,'' \emph{The Amer.
  Statist.}, vol.~58, no.~1, pp. 30--37, 2004.

\bibitem{sun2016majorization}
Y.~Sun, P.~Babu, and D.~P. Palomar, ``Majorization-minimization algorithms in
  signal processing, communications, and machine learning,'' \emph{IEEE Trans.
  Signal Process.}, vol.~65, no.~3, pp. 794--816, 2016.

\bibitem{shen2018fractional2}
K.~Shen and W.~Yu, ``Fractional programming for communication systems---{Part
  II}: {Uplink} scheduling via matching,'' \emph{IEEE Trans. Signal Process.},
  vol.~66, no.~10, pp. 2631--2644, 2018.

\bibitem{shi2015secure}
Q.~Shi, W.~Xu, J.~Wu, E.~Song, and Y.~Wang, ``Secure beamforming for {MIMO}
  broadcasting with wireless information and power transfer,'' \emph{IEEE
  Trans. Wireless Commun.}, vol.~14, no.~5, pp. 2841--2853, 2015.

\bibitem{bertsekas1999nonlinear}
D.~P. Bertsekas, \emph{Nonlinear programming}.\hskip 1em plus 0.5em minus
  0.4em\relax Belmont, MA, USA: Athena Sci., 1999.

\bibitem{BSUM}
M.~Razaviyayn, M.~Hong, and Z.-Q. Luo, ``A unified convergence analysis of
  block successive minimization methods for nonsmooth optimization,''
  \emph{SIAM J. Opt.}, vol.~23, no.~2, pp. 1126--1153, 2013.

\bibitem{phan2023inertial}
D.~N. Phan and N.~Gillis, ``An inertial block majorization minimization
  framework for nonsmooth nonconvex optimization,'' \emph{J. Mach. Learn.
  Res.}, vol.~24, pp. 1--41, 2023.

\bibitem{grippo2000convergence}
L.~Grippo and M.~Sciandrone, ``On the convergence of the block nonlinear
  gauss--seidel method under convex constraints,'' \emph{Oper. Res. Lett.},
  vol.~26, no.~3, pp. 127--136, 2000.

\bibitem{tseng2001convergence}
P.~Tseng, ``Convergence of a block coordinate descent method for
  nondifferentiable minimization,'' \emph{J. Opt. Theory and Appl.}, vol. 109,
  no.~3, p. 475, 2001.

\bibitem{higham2002accuracy}
N.~J. Higham, \emph{Accuracy and stability of numerical algorithms}.\hskip 1em
  plus 0.5em minus 0.4em\relax SIAM, 2002.

\bibitem{yu2007transmitter}
W.~Yu and T.~Lan, ``Transmitter optimization for the multi-antenna downlink
  with per-antenna power constraints,'' \emph{IEEE Trans. Signal Process.},
  vol.~55, no.~6, pp. 2646--2660, 2007.

\bibitem{hjorungnes2011complex}
A.~Hj{\o}rungnes, \emph{Complex-valued matrix derivatives: with applications in
  signal processing and communications}.\hskip 1em plus 0.5em minus 0.4em\relax
  Cambridge Uni. Press, 2011.

\bibitem{welling2010kalman}
M.~Welling, ``The {Kalman} filter,'' \emph{Lecture Note}, pp. 92--117, 2010.

\bibitem{monga2021algorithm}
V.~Monga, Y.~Li, and Y.~C. Eldar, ``Algorithm unrolling: Interpretable,
  efficient deep learning for signal and image processing,'' \emph{IEEE Signal
  Process. Mag.}, vol.~38, no.~2, pp. 18--44, 2021.

\end{thebibliography}

\end{document}